\documentclass[]{article}
\usepackage[englisch,randbleibt,abstandbleibt,nummganz]{Arbeit}

\usepackage{algpseudocode}
\usepackage{pstricks,pst-node}
\newcommand{\PDLtree}{$\mKlasse{PDL}_{\text{tree}}$}
\newcommand{\Left}{\texttt{left}}
\newcommand{\Right}{\texttt{right}}
\newcommand{\Up}{\texttt{up}}
\newcommand{\Down}{\texttt{down}}
\newcommand{\Dn}{\texttt{dn}}

\newcommand{\hld}{$\mKlasse{HL}^\downarrow$}
\newcommand{\mybox}{\Box}
\newcommand{\mydiamond}{\Diamond}
\newcommand{\NEXP}{NEXPTIME}
\newcommand{\EXP}{EXPTIME}
\newcommand{\DEXP}{2EXPTIME}
\newcommand{\NEXPTIME}{\KK{\NEXP}}
\newcommand{\mce}{\ensuremath{\text{\sffamily MC}_=}}

\newcounter{Tab}
\date{June 24, 2008}

\begin{document}

\title{Complexity of Hybrid Logics over \mbox{Transitive Frames}\footnote{This work was presented at 
              \emph{Methods for Modalities 4} (2005).}}

\author{Martin {Mundhenk}\thanks{{Institut f\"ur Informatik, Friedrich-Schiller-Universit\"at Jena, Germany}, {\texttt{mundhenk@cs.uni-jena.de}}}
        \and
        Thomas {Schneider}\thanks{School of Computer Science, University of Manchester, UK, \texttt{schneider@cs.uni-jena.de}}
        \and
        Thomas {Schwentick}\thanks{Fachbereich Informatik, Universit\"at Dortmund, Germany, \texttt{thomas.schwentick@udo.edu}} 
        \and
        Volker {Weber}\thanks{Fachbereich Informatik, Universit\"at Dortmund, Germany, \texttt{volker.weber@udo.edu}}%
}

\maketitle

  \begin{abstract}
    This article examines the complexity of hybrid logics over transitive frames,
    transitive trees, and linear frames.

    We show that satisfiability over transitive frames for the hybrid language
    extended with the downarrow operator $\dnA$ is \compl{\NEXP}. This is in contrast to undecidability
    of satisfiability over arbitrary frames for this language~\cite{ABM99}.
    It is also shown that adding the \at\ operator or the past modality leads to undecidability
    over transitive frames.

    This is again in contrast to the case of transitive trees and linear frames, 
    where we show these languages to be nonelementarily decidable.

    Moreover, we establish \KK{\DEXP} and \KK{\EXP} upper bounds for satisfiability over
    transitive frames and transitive trees, respectively, for the hybrid
    \Mop{Until}/\Mop{Since} language.
    An \KK{\EXP} lower bound is shown to hold for the modal \Mop{Until} language over both frame
    classes.
  \end{abstract}

\section{Introduction}
  \label{sec:Intro}

  Hybrid languages are extensions of  modal logic that allow for
  naming and accessing states of a model explicitly. This renders hybrid logic an adequate
  representation formalism for many applications, where the basic modal and/or temporal languages
  do not suffice. Moreover, reasoning systems are easier to devise for hybrid than for
  modal logic.

  \medskip\noindent
  \textbf{Hybrid Logic},~
  as well as the foundations of temporal logic, goes back to Arthur
  Prior~\cite{Pri67}. Since then, many\,---\,more or less powerful\,---\,languages
  have been studied.
  Here we briefly introduce the extensions that shall concern us in this article.

  \textit{Nominals} are special atomic formulae that name states of models.
  They allow, for instance, for an axiom expressing irreflexivity, which cannot be
  captured by modal formulae: $i \impl \neg\Diamond i$.

  \textit{The at operator} \at\ can be used to directly jump to states named by nominals,
  independently of the accessibility relation. Hence, the above
  formula could also be written as $\at_i \neg\Diamond i$.

  With the help of \textit{the downarrow operator $\dnA$}\,, it is possible to
  bind variables to states. Whenever $\dna x$ is encountered during the evaluation of a
  formula, the variable $x$ is bound to the current state $s$. All occurrences of $x$
  in the scope of this $\dnA$ are treated like nominals naming $s$. As an example, the formula
  $\dna x.\neg\Diamond\Diamond x$ reads as: Name the current state $x$ and make sure that
  it is not possible to go from $x$ to $x$ in exactly two steps. This is an axiom for asymmetry,
  another property not expressible in modal logic.

  Combined with the \at\ operator, $\dnA$ leads to a very powerful language that can formulate
  many desirable properties and goes far beyond the scope of the simple nominal language.
  To give a more impressive example, we consider the \Mop{Until} operator.
  The formula $\Unt(\varphi,\psi)$ reads as ``there is a point
  in the future at which $\varphi$ holds, and at all points between now and this point,
  $\psi$ holds''. What the basic modal language is not able to express, can be achieved by
  the hybrid $\dnA$-\at\ language.
  \[
      \Unt(\varphi,\psi) \equiv \Dna x.\Diamond\!\dna y.\varphi \und \at_x\Box(\Diamond y \impl \psi).
  \]

  Besides more advanced temporal concepts such as ``until'' or ``since'', hybrid
  temporal languages can express other desirable temporal notions such as ``now'',
  ``yesterday'', ``today'', or ``tomorrow''. Moreover, with hybrid logic one can
  capture many temporally relevant frame properties (besides the above mentioned,
  antisymmetry, trichotomy, directedness, \dots). For this reason, hybrid
  temporal languages are of great interest where basic temporal logic reaches its
  limits \cite{BT99,Bla00,ABM01,FdRS03}.

  \medskip\noindent
  \textbf{Transitive Frames.}~
  Hybrid logic is interpreted over Kripke frames and models, as is modal logic.
  A frame consists of a set of states (points in time) and an accessibility 
  relation $R$, where $xRy$ says that $y$ is reachable from $x$ or, seen 
  temporally, $y$ is in the future of $x$.
  We examine the computational complexity of satisfiability for several hybrid logics over
  transitive frames, transitive trees and linear frames.

  Modal, hybrid, and first-order logics over transitive models have been studied recently
  in \cite{ABM00,GMV99,ST01,Kie02,Kie03,IRR+04,DO05}. Although the complexity of
  hybrid (tense) logic has been extensively examined
  \cite{BS95,Gor96,ABM99,ABM00,FdRS03}, there are highly expressive hybrid languages
  for whose satisfiability problems only results over arbitrary, but not over
  restricted, temporally relevant frame classes have been known.

  We concentrate on transitive frames because transitivity is a property that the 
  relations of many different temporal applications have in common, even if they differ in
  other properties such as tree-likeness, trichotomy, irreflexivity, or asymmetry.
  Transitivity can be seen as the minimal requirement in many applications,
  for example temporal verification.

  But there are other reasons why this frame class is of interest, particularly in
  connection with computational complexity.
  In the special case of linear frames, nominals and \at\ can be simulated using the
  conventional modal operator and its converse. Hence, the basic hybrid language is
  as expressive over linear frames as the basic modal language. The $\dnA$ operator is
  useless even on transitive trees, a representation of branching time.
  Over transitive frames, in contrast, these hybrid operators \textit{do} make a difference. In this case, there are
  properties that can be expressed in the hybrid, but not in the modal language (see
  the irreflexivity example above). For this reason, the class of transitive frames
  can be regarded as a restricted frame class that is still general enough to separate
  hybrid from modal languages in terms of expressive power.

  Yet another reason for considering precisely transitive frames will become clear in
  the next paragraph.

  \medskip\noindent
  \textbf{Complexity of Hybrid Logics.}~
  We use complexity classes \KK{NP}, \KK{PSPACE}, \KK{\EXP},
  \KK{\NEXP},
  \KK{$n$\EXP},
  $n \ge 2$,
  and \KK{coRE} as known from~\cite{Pap94}. A problem is nonelementarily decidable
  if it is decidable but not contained in any \KK{$n$\EXP}.

  It goes without saying that reasoning tasks for
  richer logics require more resources than those for simpler languages, such as the basic
  modal language. We focus on one reasoning task, namely satisfiability. The
  modal and temporal satisfiability problems over arbitrary as well as over transitive frames are
  \compl{PSPACE}~\cite{Lad77,Spa93b}. If the ``somewhere'' modality $\Exi$ is added,
  satisfiability becomes \compl{\EXP} over arbitrary frames~\cite{Spa93a}.
  For many, more restricted, frame classes, modal and temporal satisfiability is
  \compl{NP}~\cite{Lad77,ON80,SC85}. In contrast, the known part of the complexity spectrum
  of hybrid satisfiability reaches up to undecidability.

  Many complexity results for hybrid languages have been established in
  \cite{ABM99,ABM00}. It was proven in~\cite{ABM99} that the hybrid language
  with nominals and \at\ has a \compl{PSPACE} satisfiability problem and
  that satisfiability for the hybrid tense language is \compl{\EXP}, even
  if \at\ or \Mop{E} are added.
  The same authors show that these problems have the same complexity
  (or drop to \compl{PSPACE} or \compl{NP}, respectively) if the class of frames
  is restricted to transitive frames (or transitive trees, or linear frames,
  respectively) \cite{ABM00}.

  Moreover, they established \Compl{\EXP} of satisfiability
  for the hybrid \Mop{Until}/\Mop{Since} language. The complexity of this language
  over transitive frames and transitive trees, respectively, has been open.
  \Compl{PSPACE} over linear frames is known from~\cite{FdRS03}.
  We want to find out at which exact require\-ments to the frame classes the
  decrease from \KK{\EXP} to \KK{PSPACE} takes place.

  Undecidability results for languages containing $\dnA$ originate from
  \cite{BS95,Gor96}. The strongest such result, namely for the pure nominal-free fragment
  of the \mbox{$\dnA$ language}, is given in \cite{ABM99}.
  
  In recent work~\cite{tCF05}, it was demonstrated that decidability of the $\dnA$ language
  can be regained by certain restrictions on the frame classes. Transitivity might be
  another property under which the $\dnA$ language can be ``tamed'', since it has already 
  been observed that over transitive trees and linear orders, the $\dnA$ operator on 
  its own is useless.

  \medskip\noindent
  \textbf{New Road-Map Pages.}~
  This article establishes two groups of complexity results for hybrid languages
  over transitive frames, transitive trees, and linear frames.

  First, we examine satisfiability of the hybrid \mbox{$\dnA$ language}. Our most surprising
  result is the ``taming'' of this language over transitive frames: the satisfiability problem
  is \compl{\NEXP}. This high level of complexity is retained even over complete frames. 
  We also show that enriching the language
  by the backward-looking modality $\Pas$ or the \at\ operator leads to undecidability in the
  case of transitive frames. 
  
  The situation is different over transitive trees. Decidability, even for the richest $\dnA$ 
  language, is easy to see, but we will show it to be nonelementary if $\Pas$ or \at\ are added.
  For linear frames, this is already known in the temporal case. We prove that adding \at\ suffices to
  obtain nonelementary complexity.

  As a second step, we consider satisfiability over transitive frames and transitive trees for the hybrid
  \Mop{Until}/\Mop{Since}-$\Exi$ language. We establish \Hard{\EXP} for the
  modal language extended with \Mop{Until} only. This is matched by an \KK{\EXP} upper bound for
  the full language in the case of transitive trees. As for transitive frames, we give a
  \KK{\DEXP} upper bound.

  Table~\ref{tab:Resultate} gives an overview of the satisfiability problems considered
  in this article (marked bold) and visualizes how our results arrange into a
  collection of previously known results. It makes use of the notation of hybrid languages
  introduced in Section~\ref{sec:ML_HL}.
  Complexity classes without addition stand for completeness results; ``nonel.'' stands for
  nonelementarily decidable.
  The work from which the results originate, is cited.
  Conclusions from surrounding results are abbreviated by ``c.''.
  
    \newlength{\ab}
    \setlength{\ab}{2pt}
    \newcommand{\stab}{\rule{0pt}{9pt}}
    \newcommand{\Stab}{\rule{0pt}{11pt}}
    \newcommand{\staB}{\rule[-6pt]{0pt}{4pt}}
    \newcommand{\Ml}{\mKlasse{ML}}
    \newcommand{\Hl}{\mKlasse{HL}}
    \newcommand{\NP}{\KK{NP}}
    \newcommand{\PS}{\KK{PSPACE}}
    \newcommand{\EX}{\KK{\EXP}}
    \newcommand{\Un}{coRE}
    \newcommand{\No}{nonel.}
    \newcommand{\eigETT}[2]{\textbf{\EXP} (\ref{theo:#1},\ref{theo:#2})}
    \newcommand{\eigET}[1]{\textbf{\EXP-hard} (\ref{theo:#1})}
    \newcommand{\eigDET}[1]{\textbf{in \DEXP} (\ref{theo:#1})}
    \newcommand{\eigNET}[1]{\textbf{\NEXP} (\ref{#1})}
    \newcommand{\eigUT}[1]{\textbf{\Un} (\ref{theo:#1})}
    \newcommand{\eignon}[1]{\textbf{\No} (\ref{#1})}
    {\small

  \begin{center}
      \begin{tabular}{@{~}l|l|l|l|l@{~}}
        hybrid                       & complexity    & complexity over    & complexity over          & complexity             \\
        lang.                        & over arbi-    & transitive frames  & transitive trees         & over linear            \\
                                     & trary frames  &                    &                          & orders                 \\
        \hline\Stab
        $\Hl^\at$                    & \PS\ \cite{ABM99}      & \PS\ \cite{ABM00}          & \PS\ \cite{ABM00}            & \NP\ \cite{ABM00}          \\[\ab]
        $\Hl_{\Fut,\Pas}$            & \EX\ \cite{ABM99}      & \EX\ \cite{ABM00}           & \PS\ \cite{ABM00}                 & \NP\ \cite{ABM00}              \\[\ab]
        $\Hl^\Exi_{\Fut,\Pas}$       & \EX\ \cite{ABM00}      & \EX\ \cite{ABM00}           & \PS\ \cite{ABM00}                 & \NP\ \cite{ABM00}               \\[\ab]
        $\Hl^\Exi_{\Unt,\Sin}$       & \EX\ \cite{ABM00}      & \eigDET{US_ub},    & \eigETT{US_lb}{US_tt_ub} & \PS-                   \\[\ab]
                                     &               & \eigET{US_lb}      &                          &  hard \cite{Rey03}              \\[\ab]
        \hline\Stab
        $\Hl^\dna$                   & \Un\ \cite{ABM99}      & \eigNET{ThHLDSAT}  & \PS\ \cite{ABM00}                 & \NP\ \cite{FdRS03}               \\[\ab]
        $\Hl^{\dnA,\at}$             & \Un\ \cite{ABM99}      & \eigUT{dna_at_lb}  & \eignon{theo:dna_nonel}  & \eignon{ThHLDSATlin}  \\[\ab]
        $\Hl_{\Fut,\Pas}^\dna$       & \Un\ \cite{ABM99} & \eigUT{dna_P_lb}   & \eignon{theo:dna_nonel}  & \No\ \cite{FdRS03}          \\[\ab]
        \staB
        $\Hl_{\Fut,\Pas}^{\dnA,\at}$ & \Un\ \cite{ABM99} & \textbf{\Un} (c.)  & \eignon{theo:dna_nonel}  & \No\ \cite{FdRS03}              \\
      \end{tabular}
    \end{center}
      \begin{footnotesize}
      \refstepcounter{Tab}\label{tab:Resultate}
      \emph{Table~\theTab.} An overview of complexity results for hybrid logics. Numbers in round parentheses refer to
      the corresponding theorem. 
      \end{footnotesize}\medskip
    }

  \medskip\noindent
  \textbf{Legend.}~ This article is organized as follows. In Section~\ref{sec:ML_HL}, we give all
  necessary definitions and notations of modal and hybrid logic. We present the decidability and undecidability results
  for the hybrid $\dnA$ languages in Sections~\ref{sec:dec} and~\ref{sec:undec}.
  The hybrid \Mop{Until}/\Mop{Since} language is examined in Section~\ref{sec:US-logics_trans}.
  Section~\ref{Conclusion} contains some concluding remarks.

  \section{Modal and Hybrid Logic}
    \label{sec:ML_HL}

    We define the basic concepts and notations of modal and hybrid logic that are
    relevant for our work. The fundamentals of modal logic can be found in
    \cite{BdRV01}; those of hybrid logic in \cite{ABM99,Bla00}.

    \medskip\noindent
    \textbf{Modal Logic.}~ Let $\PROP$ be a countable set of \textit{propositional atoms}.
    The language \Klasse{ML} of modal logic is the set of all formulae of the form
    \[
        \varphi ::= p \mid \neg\varphi \mid \varphi\und\varphi' \mid \Diamond\varphi\,,
    \]
    where $p \in \PROP$.
    We use the well-known abbreviations $\oder$, $\impl$, $\aequ$, $\top$ (``true''),
    and $\bot$ (``false''), as well as
    $\Box\varphi := \neg\Diamond\neg\varphi$.

    The semantics are defined via \textit{Kripke models}. Such a model is a triple $\mKlasse{M} = (M,R,V)$,
    where $M$ is a nonempty set of \textit{states}, $R \subseteq M \times M$ is a binary
    relation\,---\,the \textit{accessibility relation}\,\mbox{---\,,} and $V: \PROP \to \PM(M)$
    is a function\,---\,the \textit{valuation function}. The structure
    $\mKlasse{F} = (M,R)$ is called a \textit{frame}.
    Given a model $\mKlasse{M} = (M,R,V)$ and a state $m \in M$, the
    \textit{satisfaction relation} is defined by
    \begin{alignat*}{2}
      \mKlasse{M},m & \wahr p               & & \text{~~iff~~} m \in V(p),~ p\in\PROP,            \\
      \mKlasse{M},m & \wahr \neg\varphi     & & \text{~~iff~~} \mKlasse{M},m \not\wahr \varphi,   \\
      \mKlasse{M},m & \wahr \varphi\und\psi & & \text{~~iff~~} \mKlasse{M},m \wahr \varphi
                                                \Und \mKlasse{M},m \wahr \psi,                    \\
      \mKlasse{M},m & \wahr \Diamond\psi    & & \text{~~iff~~} \EQ n \in M(mRn \Und \mKlasse{M},n \wahr \psi).
    \end{alignat*}

    A formula $\varphi$ is \textit{satisfiable} if there exist a model $\mKlasse{M} = (M,R,V)$
    and a state $m \in M$, such that $\mKlasse{M},m \wahr \varphi$. If all states from \Klasse{M}
    satisfy $\varphi$, we write $\mKlasse{M} \wahr \varphi$ and say that $\varphi$ is
    \textit{globally satisfied} by \Klasse{M}.

    \medskip\noindent
    \textbf{Temporal Logic.}~ The language of temporal logic (tense logic)
    is the set of all formulae of the form
    \[
        \varphi ::= p \mid \neg\varphi \mid \varphi\und\varphi' \mid \Fut\varphi
                      \mid \Pas\varphi\,,
    \]
    where $p \in \PROP$.
    It is common practice to use the abbreviations
    $\Goi\varphi := \neg\Fut\neg\varphi$ and $\Has\varphi := \neg\Pas\neg\varphi$.
    Satisfaction for $\Fut$ and $\Pas$ formulae is defined by
    \begin{alignat*}{2}
      \mKlasse{M},m & \wahr \Fut\psi    & & \text{~~iff~~} \EQ n \in M(mRn \Und \mKlasse{M},n \wahr \psi), \\
      \mKlasse{M},m & \wahr \Pas\psi    & & \text{~~iff~~} \EQ n \in M(nRm \Und \mKlasse{M},n \wahr \psi).
    \end{alignat*}

    Whenever one wants to speak not only of states accessible from the current state, but
    also of states ``between'' the current and some accessible state, one can make use
    of the binary operators \Mop{U} (``until'') and \Mop{S} (``since''), for which satisfaction
    is defined by
    \begin{align*}
      \mKlasse{M},m\!\wahr\!\Unt(\varphi,\psi)
        & \text{\,~iff~\,} \EQ n \big(mRn \Und \mKlasse{M},n\!\wahr\!\varphi
        \Und \AQ s (mRsRn \Impl \mKlasse{M},s\!\wahr\!\psi)\big),\\
      \mKlasse{M},m\!\wahr\!\Sin(\varphi,\psi)
        & \text{\,~iff~\,} \EQ n \big(nRm \Und \mKlasse{M},n\!\wahr\!\varphi
        \Und \AQ s (nRsRm \Impl \mKlasse{M},s\!\wahr\!\psi)\big).
    \end{align*}
    The \Mop{U}/\Mop{S} language is strictly stronger than
    the basic temporal language in the sense that \Mop{F} and \Mop{P} can be expressed by \Mop{U}
    and \Mop{S} $\big(\text{e.\,g.}~ \Fut\varphi = \Unt(\varphi,\top)\big)$, but not vice versa.

    In~\cite{ABM00}, a variant of the \Mop{U}/\Mop{S} operators, $\Unt^{+}$ and $\Sin^{+}$,
    is introduced. Satisfaction for $\Unt^{+}$ (analogously for $\Sin^{+}$) is defined by
    \begin{align*}
      & \mKlasse{M},m \wahr \Unt^+(\varphi,\psi)                                            \\
                    & \text{~iff~~} \EQ n \in M \big(mRn \Und \mKlasse{M},n \wahr \varphi
                    \Und \AQ s \in M (mR^+sR^+n \Impl \mKlasse{M},s \wahr \psi)\big),
    \end{align*}
    where $R^+$ is the transitive closure of $R$. By means of these operators,
    they ``simulated'' transitive frames syntactically \cite{ABM00}. 

    We go a step further
    and define another modification, $\Unt^{++}$ and $\Sin^{++}$, with the satisfaction relation
    \begin{align*}
      & \mKlasse{M},m \wahr \Unt^{++}(\varphi,\psi)                                            \\
                    & \text{~iff~~} \EQ n\!\in\!M \big(m{\text{\boldmath$R^+$}}n \Und \mKlasse{M},n\!\!\;\wahr\!\!\;\varphi
                    \Und \AQ s \in M (mR^+sR^+n \Impl \mKlasse{M},s\!\!\;\wahr\!\!\;\psi)\big),
    \end{align*}
    and analogously for $\Sin^{++}$.
    The resulting temporal language is an even closer simulation of transitivity, as we will see
    in Section~\ref{sec:US-logics_trans}.

    \medskip\noindent
    \textbf{Hybrid Logic.}~
    As indicated in the previous section, \textit{the} hybrid language
    does not exist. Rather there are several extensions of the modal language allowing for
    explicit references to states and therefore being called hybrid. We introduce 
    those hybrid languages that will interest us in this article. The definitions and
    notations are taken from \cite{ABM99,ABM00}.

    Let $\NOM$ be a countable set of \textit{nominals}, $\SVAR$ be a countable set of
    \textit{state variables}, and $\ATOM = \PROP \cup \NOM \cup \SVAR$. It is common practice to write
    propositional atoms as $p,q,\dots$, nominals as $i,j,\dots$, and state variables as $x,y,\dots$
    The \textit{full hybrid language} $\mKlasse{HL}^{\dnA,\at}$ is the set of all formulae of
    the form
    \[
        \varphi ::= a \mid \neg\varphi \mid \varphi\und\varphi' \mid \Diamond\varphi
                    \mid \at_t\varphi \mid \Dna x.\varphi\,,
    \]
    where $a \in \ATOM$, $t \in \NOM \cup \SVAR$, and $x \in \SVAR$.

    A hybrid formula is called
    \textit{pure} if it contains no propositional atoms;
    \textit{nominal-free} if it contains no nominals; and
    a \textit{sentence} if it contains no free state variables. (\textit{Free} and
    \textit{bound} are defined as usual; the only binding operator here is $\dnA$\,.)

    A \textit{hybrid model} is a Kripke model with the valuation function
    $V$ extended to $\PROP \cup \NOM$, where for all $i \in \NOM$,~
    $|V(i)| = 1$. Whenever it is clear from the context, we will omit the word ``hybrid''
    when referring to models.
    In order to evaluate $\dnA$-formulae, an \textit{assignment} $g: \SVAR \to M$
    \textit{for \Klasse{M}} is necessary. Given an assignment $g$, a state variable $x$ and
    a state $m$, \textit{an $x$-variant $g^x_m$ of $g$} is defined by
    \[
        g^x_m(x') = \begin{cases}
                      m     & \text{if}~x' = x, \\
                      g(x') & \text{otherwise}.
                   \end{cases}
    \]
    For any atom $a$, let
    \[
        [V,g](a) = \begin{cases}
                     \{g(a)\} & \text{if}~a \in \SVAR, \\
                     V(a)     & \text{otherwise}.
                   \end{cases}
    \]

    The satisfaction relation for
    hybrid formulae is defined by
    \begin{alignat*}{2}
      \mKlasse{M},g,m & \wahr a               & & \text{~~iff~~} m \in [V,g](a),~ a\in\ATOM,           \\
      \mKlasse{M},g,m & \wahr \neg\varphi     & & \text{~~iff~~} \mKlasse{M},g,m \not\wahr \varphi,    \\
      \mKlasse{M},g,m & \wahr \varphi\und\psi & & \text{~~iff~~} \mKlasse{M},g,m \wahr \varphi
                                                  \Und \mKlasse{M},g,m \wahr \psi,                     \\
      \mKlasse{M},g,m & \wahr \Diamond\varphi & & \text{~~iff~~} \EQ n \in M
                                                  (mRn \Und \mKlasse{M},g,n \wahr \varphi),            \\
      \mKlasse{M},g,m & \wahr \at_t\varphi    & & \text{~~iff~~} \EQ n \in M
                                                  (\mKlasse{M},g,n\!\wahr\!\varphi\Und [V,g](t)\!=\!\{n\}),  \\
      \mKlasse{M},g,m & \wahr \Dna x.\varphi  & & \text{~~iff~~} \mKlasse{M},g^x_m,m \wahr \varphi.
    \end{alignat*}

    A formula is \textit{satisfiable} if there exist a model $\mKlasse{M} = (M,R,V)$,
    an assignment $g$ for \Klasse{M}, and a state $m \in M$, such that $\mKlasse{M},g,m \wahr \varphi$.

    We sometimes use the ``somewhere'' modality $\Exi$ having the interpretation
    \[
        \mKlasse{M},g,m \wahr \Exi\varphi \text{~~iff~~} \EQ n \in M (\mKlasse{M},g,n \wahr \varphi).
    \]
    In this case, \at\ is needless, because $\at_t\,\varphi$ can be expressed
    by $\Exi(t \und \varphi)$.

    \medskip\noindent
    \textbf{First-Order Logic.}~
    Modal and hybrid logic can be embedded into fragments of first-order logic.
    We will always use the standard notation of first-order logic.

    We will make use of certain fragments of first-order logic and denote them in the style
    of~\cite{BGG97}: $[\text{all},(u,1)]$, where $u \in \omega$. This notation stands for
    the fragment without equality, without function symbols, and with no other relation symbols
    than one binary and $u$ unary ones\footnote{Although constants are not mentioned in this notation, 
    they are always assumed present, since they correspond to nominals in hybrid logic.}.    
    We denote the satisfiability problem for such a fragment
    by $[\text{all},(u,1)]$\Probl{-SAT} and $[\text{all},(u,1)]$\Probl{-trans-SAT}, where the latter
    requires that the binary relation symbol is interpreted by a transitive relation.

    The \textit{Standard Translation} $\ST$ \cite{tCF05} embeds hybrid logic into first-order logic 
    and consists of two functions $\ST_x$ and $\ST_y$, defined recursively. Since $\ST_y$ is 
    obtained from $\ST_x$ by exchanging $x$ and $y$, we only give $\ST_x$ here.
    \begin{xalignat*}{2}
      \ST_x(p)                 & ~=~ P(x),                                        &
      \ST_x(\Diamond\varphi)   & ~=~ \EQ y \big(xRy \und \ST_y(\varphi)\big),     \\
      \ST_x(t)                 & ~=~ t\!=\!x,                                     &
      \ST_x(\at_t\varphi)      & ~=~ \EQ y \big(y\!=\!t \und \ST_y(\varphi)\big), \\
      \ST_x(\neg\varphi)       & ~=~ \neg\ST_x(\varphi),                          &
      \!\!\ST_x(\dna v.\varphi)  & ~=~ \EQ v \big(x\!=\!v \und \ST_x(\varphi)\big), \\
      \ST_x(\varphi\!\!\;\und\!\!\;\psi) & ~=~ \ST_x(\varphi) \und \ST_x(\psi),             &
      \ST_x(\Exi\varphi)       & ~=~ \EQ y \big(\ST_y(\varphi)\big),
    \end{xalignat*}
    where $p \in \PROP$, $t \in \NOM \cup \SVAR$, and $v \in \SVAR$.

    \medskip\noindent
    \textbf{Properties of Models and Frames.}~
    Let $\mKlasse{M} = (M,R,V)$ be a (Kripke or hybrid) model with the underlying frame $\mKlasse{F} = (M,R)$.
    By $R^+$ we denote the transitive closure of $R$. For any subset \mbox{$M' \subseteq M$}, we write
    \mbox{$R\einschr_{M'}$} and \mbox{$V\einschr_{M'}$} for the restrictions of $R$ and $V$ to $M'$. We will
    refer to \textit{transitive frames} or \textit{linear frames}
    whenever we mean frames whose accessibility relation is transitive or a linear order, respectively.
    A \textit{linear order} is an irreflexive, transitive, and trichotomous relation, where trichotomy
    is defined by
    \[
        \big(\AQ xy(xRy \text{~or~} x\!=\!y \text{~or~} yRx)\big).
    \]
    A frame \Klasse{F} is a \textit{tree}, if and only if it is acyclic and connected, and every point has
    at most one $R$-predecessor. A \textit{transitive tree} is any $(M,R^+)$, where $(M,R)$
    is a tree.

    \medskip\noindent
    \textbf{Satisfiability Problems.}~
    Whenever we leave one or more operators out of the hybrid language, we omit the according
    superscript of \Klasse{HL}.
    If we proceed to a hybrid tense
    language, we add the suitable temporal operator(s) as subscript(s) to \Klasse{HL}.
    Analogously, when equipping the modal language with additional operators,
    we add them as sub- or superscripts to \Klasse{ML}.

    {
    \newcommand{\oben}{x}
    \newcommand{\unten}{y}
    For any hybrid language $\mKlasse{HL}^\oben_\unten$\,, the \textit{satisfiability problem}
    $\mKlasse{HL}^\oben_\unten\Probl{-SAT}$ is defined as follows: Given a formula
    $\varphi \in \mKlasse{HL}^\oben_\unten$\,, does there exist a hybrid model \Klasse{M},
    an assignment $g$ for \Klasse{M}, and a state $m \in M$ such that
    $\mKlasse{M},g,m \wahr \varphi$\,? If $\dnA$ is not in the considered language, the
    assignment $g$ can be left out of this formulation. If we only ask for
    \textit{transitive models} (or \textit{transitive trees} or \textit{linear models}, respectively)
    satisfying $\varphi$, we speak of $\mKlasse{HL}^\oben_\unten\Probl{-trans-SAT}$
    (or $\mKlasse{HL}^\oben_\unten\Probl{-tt-SAT}$, or $\mKlasse{HL}^\oben_\unten\Probl{-lin-SAT}$, respectively).
    }
    E.g., the satisfiability problem over transitive frames
    for the hybrid temporal $\dnA$ language is denoted by
    $\mKlasse{HL}^{\dna}_{\Fut,\Pas}$-trans-SAT.

\section{Deciding $\mKlasse{HL}^\downarrow$ over Transitive Frames}
  \label{sec:dec}

Areces, Blackburn, and Marx \cite{ABM99} proved that the downarrow operator $\downarrow$ turns
the satisfiability problem for hybrid logics undecidable in general, even if no interaction with
$\at$ or $\Pas$ is allowed.

We prove that undecidability vanishes if frames are required to be transitive.
\begin{Theo}\label{ThHLDSAT}
The satisfiability problem for \hld\ over transitive frames is complete for \NEXPTIME.
\end{Theo}

Before we start with the proof, we have a first look at \hld\ over
transitive frames. Obviously, it has no finite model property. E.g., the following sentence requires a model containing an infinite chain of states labeled $p$.
$$p\wedge \mydiamond p \wedge \mybox\mydiamond p \wedge \mybox\dna x.\neg\mydiamond x$$

Neither is it always possible to find a model that is a transitive
tree. But, in some way, we can get close to this.

Although our models may contain cycles, transitivity ensures that all states in a cycle are pairwise connected. I.e., the subframe consisting of these states is complete. Therefore, we can view a model as consisting of maximal complete subframes and single states, that are connected in a transitive but acyclic fashion.

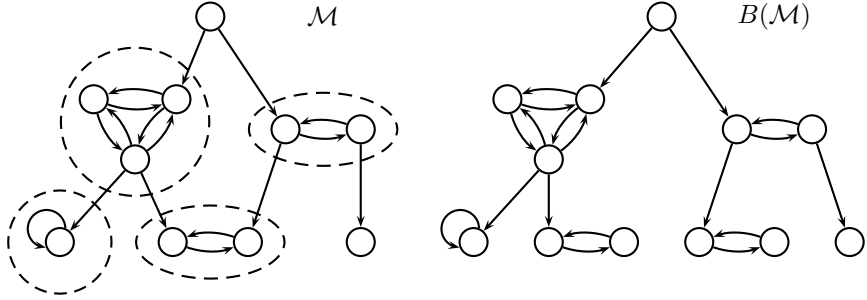
\begin{figure}
\begin{center}
\begin{pspicture}(-0.2,0)(11.2,3.8)
\rput(4,3.5){$\mKlasse{M}$}
\rput(10,3.5){$B(\mKlasse{M})$}
\cnode(2.5,3.5){2mm}{a}
\cnode(0.95,2.4){2mm}{b}
\cnode(2.05,2.4){2mm}{c}
\cnode(1.5,1.6){2mm}{d}
\cnode(3.5,2){2mm}{e}
\cnode(4.5,2){2mm}{f}
\cnode(0.5,0.5){2mm}{g}
\cnode(2,0.5){2mm}{h}
\cnode(3,0.5){2mm}{i}
\cnode(4.5,0.5){2mm}{j}
\ncline{->}{a}{c}
\ncline{->}{a}{e}
\ncline{->}{d}{g}
\ncline{->}{d}{h}
\ncline{->}{e}{i}
\ncline{->}{f}{j}
\nccircle[angle=45]{->}{g}{2.5mm}
\psset{arcangle=20}
\ncarc{<-}{b}{c}
\ncarc{<-}{b}{d}
\ncarc{<-}{d}{c}
\ncarc{<-}{e}{f}
\ncarc{<-}{h}{i}
\psset{arcangle=-20}
\ncarc{->}{b}{c}
\ncarc{->}{b}{d}
\ncarc{->}{d}{c}
\ncarc{->}{e}{f}
\ncarc{->}{h}{i}
\psset{linestyle=dashed}
\psellipse(4,2)(1,0.5)
\psellipse(1.5,2.1)(1,1)
\psellipse(2.5,0.5)(1,0.5)
\psellipse(0.5,0.5)(0.7,0.7)
\psset{linestyle=solid}
\cnode(8.5,3.5){2mm}{a2}
\cnode(6.45,2.4){2mm}{b2}
\cnode(7.55,2.4){2mm}{c2}
\cnode(7,1.6){2mm}{d2}
\cnode(9.5,2){2mm}{e2}
\cnode(10.5,2){2mm}{f2}
\cnode(6,0.5){2mm}{g2}
\cnode(7,0.5){2mm}{h2}
\cnode(8,0.5){2mm}{i2}
\cnode(9,0.5){2mm}{h3}
\cnode(10,0.5){2mm}{i3}
\cnode(11,0.5){2mm}{j2}
\ncline{->}{a2}{c2}
\ncline{->}{a2}{e2}
\ncline{->}{d2}{g2}
\ncline{->}{d2}{h2}
\ncline{->}{e2}{h3}
\ncline{->}{f2}{j2}
\nccircle[angle=45]{->}{g2}{2.5mm}
\psset{arcangle=20}
\ncarc{<-}{b2}{c2}
\ncarc{<-}{b2}{d2}
\ncarc{<-}{d2}{c2}
\ncarc{<-}{e2}{f2}
\ncarc{<-}{h2}{i2}
\ncarc{<-}{h3}{i3}
\psset{arcangle=-20}
\ncarc{->}{b2}{c2}
\ncarc{->}{b2}{d2}
\ncarc{->}{d2}{c2}
\ncarc{->}{e2}{f2}
\ncarc{->}{h2}{i2}
\ncarc{->}{h3}{i3}
\end{pspicture}
\end{center}
\caption{A transitive model $\mKlasse{M}$ and its block tree $B(\mKlasse{M})$. The maximal complete subframes of $\mKlasse{M}$ are marked by the dashed circles. Some edges caused by transitivity are left out for simplicity.}
\label{fig:blockTree}
\end{figure}

For every transitive model $\mKlasse{M} = (M,R,V)$, we define its {\em block tree} 
$B(\mKlasse{M}) = (M',R',V')$ as the structure obtained as follows.
First, we replace each maximal complete subframe of $\mKlasse{M}$, for short clique, with a single vertex.
Second, we unravel the resulting structure into a (potentially
infinite) transitive tree $T$. Then we replace each vertex of $T$ by (a
copy of) the clique of $\mKlasse{M}$ from which it is derived (Figure \ref{fig:blockTree}). 
Note that a block tree is not a tree, but we get a 
transitive tree if we view every clique as a node.

In the following, we are often interested
in this underlying tree structure of a block tree and refer to the
cliques of a block tree as {\em nodes}. For a state $s$, we denote
its node by $u_s$. We say that a node $v$ {\em is below} a node $u$ if
the states of $v$ are reachable from the states in $u$ (but not vice
versa) and a node $v$ is a {\em child} of a node $u$, if $v$ is below $u$
but there is no node $w$ below $u$ and above $v$.

Likewise, we use the terms tree, subtree,
and leaf for block tree, sub block tree, and leaf clique, respectively.

We have to be careful about how to treat nominals when unraveling
a model. If $V(i)=\{s\}$ for a nominal $i$ and a state $s\in M$, we
define $V'(i)$ to be the set of states from $M'$, that are copies via
the unraveling of $s$. Therefore, $B(\mKlasse{M})$ is not a model
\footnote{Note that we can always get a model for $\varphi$ from 
$B(\mKlasse{M})$ by joining the states labeled with the same nominals, 
but this model might be different from $\mKlasse{M}$.} as
defined in Section \ref{sec:ML_HL}, because nominals may hold at more
than one state, but by viewing $i$ as a propositional atom it can be treated
as a model. The satisfaction
relation is not affected and it is therefore easy to see that this
transformation preserves satisfaction of \hld-sentences: The relation
associating each state of $\mKlasse{M}$ with every copy in
$B(\mKlasse{M})$ is a quasi-injective bisimulation\footnote{The notion
of bisimulation has to be extended by requiring states carrying the
same nominal to be related.} \cite{BS98}.
\begin{Lem}
For every transitive model $\mKlasse{M}$, every state $s$ of  $\mKlasse{M}$, every copy $s'$ of $s$ in $B(\mKlasse{M})$, and every \hld-sentence $\varphi$:~
$$\mKlasse{M},s\models\varphi  \iff B(\mKlasse{M}),s'\models\varphi.$$
\end{Lem}

If, for some block tree $\mKlasse{B}$ and some state $s$ of $\mKlasse{B}$, $\mKlasse{B},s\models\varphi$ holds, we refer to $\mKlasse{B}$ as a block tree model for $\varphi$.

Before we show how to use this tree-like structure to decide \hld\ over transitive frames, we focus on complete subframes and show that their size can be bounded.

\subsection{\hld\ over Complete Frames}

As complete subframes are a significant part of transitive models, 
we are now going to study the satisfiability problem of
\hld\ over complete frames\footnote{All results in this subsection hold for $\mKlasse{HL}^{\dna,\at}_{\Fut,\Pas}$, too.}. The most important result for our purpose
 is an exponential-size model property of \hld\ over
complete frames.

We want to start by giving some insight why this
property holds. In complete  frames, the accessibility relation
does not distinguish different states. Of course,  states can be told
apart if they are labeled differently by propositions. But
the number of different labelings is exponentially bounded in the size
of the formula. To use more states, we have to distinguish states
labeled equally. This can only be done by assigning names to these
states. But the number of states we can distinguish in this way is
bounded by the number of different state variables and nominals used in the sentence.

While intuition is clear, we can prove this bound by observing that
\hld\ over complete frames is equivalent to the \emph{Monadic Class
  with equality} (\mce), the fragment of first-order logic with only
unary predicates, equality, and no function symbols \cite{BGG97}.

Before we present this connection precisely, we have to make a note on
models. Every hybrid model can be viewed as a relational structure for
its first-order correspondence language. This first-order language
usually contains a binary relation to reflect the accessibility
relation. For complete frames, the accessibility relation is trivial and can
be ignored, respectively added when going from \mce\ to \hld.
\begin{Lem}\label{reductions}
There are polynomial time functions mapping \hld\ formulas $\varphi$ to
\emph{\mce} formulas $\psi$ and vice versa such that $\varphi$ holds
in a complete hybrid model $\mKlasse{M}$ if and only if $\psi$
holds in the corresponding monadic structure.
\end{Lem}
\begin{Bew}
The mapping from \hld\ over complete frames to \mce\ is based on the
Standard Translation $\ST$ as defined in Section~\ref{sec:ML_HL}.
The only rule of $\ST$ that uses the binary relation is the rule for the diamond operator:
$\text{ST}_x(\mydiamond\alpha)=\exists y(xRy\wedge \text{ST}_y(\alpha)).$
But the right side can be reduced to $\exists y(\text{ST}_y(\alpha))$, since $xRy$ always holds on complete frames.

For the other direction, we give the rules of a reduction HT from \mce\ to \hld.
\begin{xalignat*}{2}
  \text{HT}(P(x))              & ~=~ \mydiamond(x\wedge p)                                        &
  \text{HT}(x=y)                                                 & ~=~ \mydiamond(x\wedge y)     \\
  \text{HT}(\neg\varphi)       & ~=~ \neg\text{HT}(\varphi)                                     &
  \text{HT}(\varphi\wedge\psi) & ~=~ \text{HT}(\varphi)\wedge\text{HT}(\psi) \\
  \text{HT}(\exists x.\varphi) & ~=~ \mydiamond(\dna x.\text{HT}(\varphi))                          &
\end{xalignat*}

Note that both mappings can be computed in polynomial time and do
not blow up formula size.
\end{Bew}

This result allows us to transfer complexity results and model properties for \mce\ \cite{BGG97} to \hld\ over complete frames.
\begin{Theo}\label{Cor:ExpModelProp}
\hld\ over complete frames has the exponential-size model property and its satisfiability problem is complete for \NEXPTIME.
\end{Theo}

The lower bound can be transferred directly to the case of transitive frames.
\begin{Cor}\label{CorNEXPhard}
The satisfiability problem for \hld\ over transitive frames is hard for \NEXPTIME.
\end{Cor}
\begin{Bew}
We can force a transitive frame, more precisely, the subframe generated by the current state, to be complete by adding
$\dna x.\mybox\mydiamond x$.  In this way, we can give a reduction from the satisfiability problem of \mce\ by mapping a formula $\varphi$ to $(\dna x.\mybox\mydiamond x) \wedge \text{HT}(\varphi)$.
\end{Bew}

\subsection{On Transitive Frames for \hld}\label{SubSecFrames}

Let us summarize what we have seen so far. For every \hld-formula
satisfiable over transitive frames, instead of a transitive model we
can consider its block tree. The size of the
cliques in the block tree can be exponentially bounded in the size of
the formula by Theorem \ref {Cor:ExpModelProp}.

The algorithm for testing  \hld-satisfiability will essentially guess
a model and verify that it is correct. As there is no finite model
property, all models might be infinite. Nevertheless, we will show
that, if the formula is satisfiable, there is always a model with a
regular structure in which certain finite patterns are repeated
infinitely often. This will allow us to find a finite representation
of such a model.

To this end, Definition \ref{def:type} captures the information about a
state of a block tree that will be needed for the
following. Intuitively, the $\varphi$-type of a state captures the information
 needed about its subtrees in order to evaluate any subformula of a
 given formula $\varphi$ at this state.  Here, $\psi[free/\bot]$ is
 the sentence obtained from $\psi$ by replacing every free variable by
 $\bot$ and $sub(\varphi)$ is the set of all \mbox{subformulae of
 $\varphi$.}
\begin{Def}\label{def:type}
Let $\varphi$ be a \hld-sentence and $\mKlasse{B}=(M,R,V)$ a block
tree model which is a model of $\varphi$. The \emph{$\varphi$-type} of
a state $s\in M$ is the set of all sentences from $\{ \psi[free/\bot]
\;|\; \mydiamond\psi\in sub(\varphi) \}$ that hold at some state in
the subtree \mbox{rooted at $s$.}
\end{Def}

Note that states in the same clique have the same
$\varphi$-type. Therefore, we can speak of the $\varphi$-type of a
node. The type of a node is always a superset of the types of its
children. More precisely, it is always the union of the types of the
children together with the set of relevant formulae which hold in the
node itself.

When evaluating a subformula of a \hld-sentence $\varphi$ at some
state $s$ of a block tree, all we need to
know about states strictly below $u_s$ are the
$\varphi$-types of the children of $u_s$. I.e., we can replace
subtrees below $u_s$ by subtrees of the same $\varphi$-type.
In the following lemma,  for a block tree  $\mKlasse{B}$ and two
states $s_1,s_2$, $\mKlasse{B}[u_{s_1}/u_{s_2}]$ denotes
the block tree resulting from $\mKlasse{B}$ by
replacing the subtree rooted at $u_{s_1}$ by the subtree rooted at
$u_{s_2}$. The result of this substitution is again a block tree.
\begin{Lem}\label{LemSubstitution}
Let $\varphi$ be a \hld-sentence, $\mKlasse{B}=(M,R,V)$ a block tree
model of $\varphi$ and $s_1$ and $s_2$ states of $\mKlasse{M}$ such that there
is a path from $s_1$ to $s_2$ but not vice versa. For every formula
$\psi\in sub(\varphi)$, every state $s_3$ of $\mKlasse{M}$ of the
same $\varphi$-type as $s_2$, and every assignment $g$ that maps all
free variables in $\psi$ to $s_1$ or states in $M$ preceding $s_1$:
$$\mKlasse{B},g,s_1\models \psi \iff
\mKlasse{B}[u_{s_2}/u_{s_3}],g,s_1\models \psi.$$
\end{Lem}
\begin{Bew}
The proof is by induction on the structure of $\psi$. Most cases are trivial since $s_1$ is the only state that has to be considered, e.g., if $\psi=\psi_1\wedge\psi_2$ we only need to know whether $\psi_1$ and $\psi_2$ hold at $s_1$.

The interesting case is $\psi=\mydiamond\xi$. Here, we need to know
whether $\xi$ holds at some state $s'$ reachable from $s_1$. This can
only be affected by our substitution if $s'$ is in the replaced
subtree, and therefore strictly below $s_1$. Thus, by our assumption on $g$,
all free variables in $\psi$ are mapped to states different and not
reachable from $s'$. We can conclude
$$\mKlasse{B},g,s'\models\xi\iff\mathcal{B},g,s'\models\xi[free/\bot].$$
Hence $\xi$ holds at some state in the replaced subtree rooted at
$u_{s_2}$, if and only if $\xi[free/\bot]$ is in the $\varphi$-type of
$u_{s_2}$. The lemma follows because the new subtree has the same
$\varphi$-type.
\end{Bew}

Note that we restricted the choice of $g$ only to those assignments that 
are relevant when evaluating the sentence $\varphi$.

We can use the previous lemma to get some nice restrictions on the
block trees under consideration. E.g., we can assume that for every
sentence in the type of a node, there is a witness in the node itself
or in one of its children.
\begin{Lem}\label{Lem:Models}
Let $\varphi$ be a \hld-sentence satisfiable over transitive
frames. Then there is a block tree model $\mKlasse{B}$ for $\varphi$,
in which
\begin{itemize}
        \item every node has at most $|\varphi|$ children,
        \item for every node $u$ with $\varphi$-type $t$ and every \hld-sentence $\psi\in t$, $\psi$ holds at a state in $u$ or at a state in a child of $u$, and
        \item on every path from the root, infinite or ending at a leaf, every $\varphi$-type occurs only once or infinitely often.
\end{itemize}
\end{Lem}
\begin{Bew}
Let $\varphi$ be a \hld-sentence satisfiable over transitive frames
and $\mKlasse{B'}$ a block tree model for $\varphi$.

A block tree satisfying the third condition can be obtained from
$\mKlasse{B'}$ by applying Lemma \ref{LemSubstitution}. If there are
two nodes $u$ and $v$ on a path, $v$ below $u$, that have the same
$\varphi$-type, we can replace the subtree rooted at $u$ by the
subtree rooted at $v$. This allows us to cut every finite repetition
down to a single occurrence of a $\varphi$-type. The resulting
structure is still a block tree model for $\varphi$.

Now, consider some node $u$ and its $\varphi$-type $t$. For every
sentence $\psi\in t$, we select some state in the subtree rooted at $u$
such that $\psi$ holds at this state. Let $u_1,\ldots,u_k$ be the
nodes of those states, that are not in $u$. It easy to see, following
similar tracks as in Lemma
\ref{LemSubstitution}, that the block tree obtained by
removing the nodes below $u$ and inserting instead $u_1,\ldots,u_k$ as
children of $u$ is again a block tree model of $\varphi$. Even more, the type of
$u$ is not changed by this replacement.

By applying this argument top-down from the root, we get a block tree
model for $\varphi$ satisfying the first two conditions. The third one
is not affected by this transformation.

Note that some care is needed to make this approach work for infinite
models. Basically, we must define a function that assigns to each
node of the original model its set of witnesses. The resulting model is
obtained by using this function in a straightforward fashion.
\end{Bew}

\subsection{Deciding \hld-SAT over Transitive Frames}

We will now finish the proof of Theorem \ref{ThHLDSAT} by presenting a
nondeterministic algorithm that decides \hld-SAT over transitive
frames in exponential time, basically by guessing and verifying the
finite representation of a block tree model for a given \hld-sentence
$\varphi$.

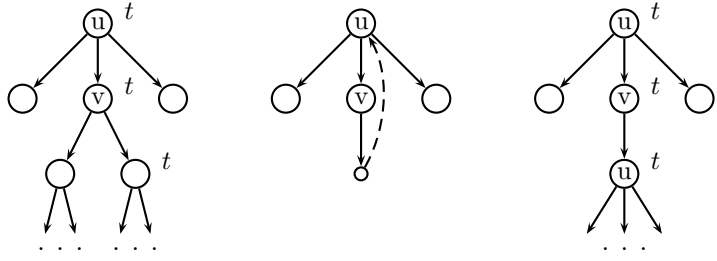
\begin{figure}
\begin{center}
\begin{pspicture}(0,0.4)(10,4)
\cnode(1.5,3.5){2mm}{a1}
\cnode(.5,2.5){2mm}{a2}
\cnode(1.5,2.5){2mm}{a3}
\cnode(2.5,2.5){2mm}{a4}
\cnode(1,1.5){2mm}{a5}
\cnode(2,1.5){2mm}{a6}
\pnode(0.8,0.7){a7}
\pnode(1.2,0.7){a8}
\pnode(1.8,0.7){a9}
\pnode(2.2,0.7){a10}
\rput(1,0.5){. . .}
\rput(2,0.5){. . .}
\rput(1.5,3.5){u}
\rput(1.5,2.5){v}
\ncline{->}{a1}{a2}
\ncline{->}{a1}{a3}
\ncline{->}{a1}{a4}
\ncline{->}{a3}{a5}
\ncline{->}{a3}{a6}
\ncline{->}{a5}{a7}
\ncline{->}{a5}{a8}
\ncline{->}{a6}{a9}
\ncline{->}{a6}{a10}
\nput{20}{a1}{$t$}
\nput{20}{a3}{$t$}
\nput{20}{a6}{$t$}

\cnode(5,3.5){2mm}{b1}
\cnode(4,2.5){2mm}{b2}
\cnode(5,2.5){2mm}{b3}
\cnode(6,2.5){2mm}{b4}
\cnode(5,1.5){1mm}{b5}
\rput(5,3.5){u}
\rput(5,2.5){v}
\ncline{->}{b1}{b2}
\ncline{->}{b1}{b3}
\ncline{->}{b1}{b4}
\ncline{->}{b3}{b5}
\ncarc[arcangle=-30,linestyle=dashed]{->}{b5}{b1}

\cnode(8.5,3.5){2mm}{c1}
\cnode(7.5,2.5){2mm}{c2}
\cnode(8.5,2.5){2mm}{c3}
\cnode(9.5,2.5){2mm}{c4}
\cnode(8.5,1.5){2mm}{c5}
\pnode(8,0.7){c6}
\pnode(8.5,0.7){c7}
\pnode(9,0.7){c8}
\rput(8.5,0.5){. . .}
\rput(8.5,3.5){u}
\rput(8.5,2.5){v}
\rput(8.5,1.5){u}
\ncline{->}{c1}{c2}
\ncline{->}{c1}{c3}
\ncline{->}{c1}{c4}
\ncline{->}{c3}{c5}
\ncline{->}{c5}{c6}
\ncline{->}{c5}{c7}
\ncline{->}{c5}{c8}
\nput{20}{c1}{$t$}
\nput{20}{c3}{$t$}
\nput{20}{c5}{$t$}

\end{pspicture}
\end{center}
\caption{An infinite block tree, its finite representation, and the infinite block tree obtained from this representation.}
\label{fig:finRep}
\end{figure}

Given a block tree $\mKlasse{B}$ with the properties of Lemma
\ref{Lem:Models}, we get a finite representation as follows. For each
path of $\mKlasse{B}$ we consider the first node $v$ that has the same
type as its parent node $u$. We replace the subtree below $v$ by a
single state labeled with a reference to $u$, as shown in Figure \ref{fig:finRep}. We need to keep $v$ because it might be the only witness for a formula in the $\varphi$-type of $u$ (cf. Lemma \ref{Lem:Models}). Clearly, the resulting
structure is finite.

By Lemma \ref{LemSubstitution} and Lemma \ref{Lem:Models}, we can get
a block tree model from this representation by replacing
each reference with the subtree rooted at the referenced node, i.e.,
essentially by an unraveling (Figure \ref{fig:finRep}).

Due  to Lemma \ref{LemSubstitution}, the size of the representation
can be reduced even further.  If there are two nodes $u$ and
$v$ of the same $\varphi$-type which are both
the first node of their type on their
path from the root, we can replace the subtree rooted at $v$ with the
subtree of $u$. I.e., whenever two nodes have the same
$\varphi$-type, we can assume that their generated subtrees are
equal. We have to check them only once.

\begin{figure}
\begin{center}
\begin{pspicture}(0,0.4)(8,3)
\cnode(2,2.5){2mm}{a1}
\cnode(1,1.5){2mm}{a2}
\cnode(3,1.5){2mm}{a3}
\pnode(0.5,0.7){a4}
\pnode(1.5,0.7){a5}
\pnode(2.5,0.7){a6}
\pnode(3.5,0.7){a7}
\rput(1,0.5){. . .}
\rput(3,0.5){. . .}
\rput(1,1.5){u}
\rput(3,1.5){v}
\ncline{->}{a1}{a2}
\ncline{->}{a1}{a3}
\ncline{->}{a2}{a4}
\ncline{->}{a2}{a5}
\ncline{->}{a3}{a6}
\ncline{->}{a3}{a7}
\nput{160}{a2}{$t$}
\nput{20}{a3}{$t$}

\cnode(6,2.5){2mm}{b1}
\cnode(5,1.5){2mm}{b2}
\cnode(7,1.5){1mm}{b3}
\pnode(4.5,0.7){b4}
\pnode(5.5,0.7){b5}
\rput(5,0.5){. . .}
\rput(5,1.5){u}
\ncline{->}{b1}{b2}
\ncline{->}{b1}{b3}
\ncline{->}{b2}{b4}
\ncline{->}{b2}{b5}
\nput{160}{b2}{$t$}
\ncline[linestyle=dashed]{->}{b3}{b2}
\end{pspicture}
\end{center}
\caption{Elimination of duplicates in the representation of a block tree.}
\label{fig:dupElim}
\end{figure}
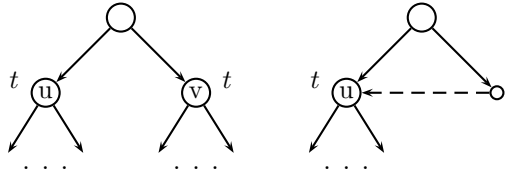

This observation can be reflected in our representation by replacing
every duplicate with a reference, as illustrated in Figure \ref{fig:dupElim}. This
causes every type to appear at most twice in the representation, thus
the number of nodes is at most exponential.

Such a representation can be described by a structure $(M\dot{\cup}C,R,V)$ such
that the states in $C$ have no outgoing edges, and a function $f$ from
$C$ to $M$. A state $s\in C$ stands for a repetition respectively duplication of the
subtree rooted at $f(s)$, including states from $C$. This causes
infinite repetition if $s$ is below $f(s)$. Note that a state in $C$ 
is a node of its own, in fact a leaf, and cannot be in the same complete subframe as a state in $M$. 

Summing up, if $\varphi$ is satisfiable, there is a representation
of a block tree model for $\varphi$ of size at most exponential in the length of
$\varphi$. The first step of the algorithm presented in Figure \ref{fig:alg} is to guess
such a representation (step 1).

In order to obtain an algorithm which tests whether the
representations indeed represents a model of $\varphi$, we describe
how to modify the
model checking algorithm MCFULL by Franceschet and de Rijke \cite{FdR05} to do so.

First, we  deal with the states in $C$ (step 2), simply by guessing their $\varphi$-types.
Next, the model checking algorithm MCFULL is used on the states in $M$ (step 3). We
 have to modify this algorithm in two respects. First, it has to use
 the information guessed for the states in $C$. Second, it should
 compute the $\varphi$-types of the states in $M$. To this end,
 it first evaluates the sentences resulting from subformulae of
 $\varphi$ by replacing free variables with $\bot$. We call this
 modified algorithm MCFULL'. The changes are straightforward.

After running MCFULL' the algorithm has computed for each state the
set of formulae that hold at this state. These sets depend on the
guesses in Step 2. Therefore, the algorithm has to verify  their consistency.
This can be done by comparing the $\varphi$-types of
the states in $C$ with the $\varphi$-types of the referenced states (steps 4 and 5).

Finally, the algorithm checks if $\varphi$ holds at some state in $M$ (step 6).

\begin{figure}
\hld-SAT$(\varphi)$
\begin{algorithmic}[1]
\State Guess the finite representation $(M\dot{\cup}C,R,V,f)$ of a block tree model for $\varphi$.
\State Guess a $\varphi$-type for every state in $C$.
\State Run MCFULL' on the states in $M$.
\State Compute the $\varphi$-type for every state in $M$ referenced by a state in $C$ via $f$.
\State Check for every state $s\in C$ that $f(s)$ has the same $\varphi$-type as $s$. If not, reject.
\State Accept iff $\varphi$ holds at some state in $M$.
\end{algorithmic}
\caption{Our algorithm for \hld-satisfiability.}
\label{fig:alg}
\end{figure}

\begin{Theo}\label{ThAlg}
The algorithm \hld-SAT presented in Figure \ref{fig:alg} decides \hld-satisfiability over transitive frames nondeterministically in exponential time.
\end{Theo}
\begin{Bew}
In Section \ref{SubSecFrames}, we have seen that for every satisfiable
sentence $\varphi$ there is a block tree model as described in
Lemma \ref{Lem:Models}. We have also seen how to represent this
block tree in a finite manner. The algorithm can guess this
representation and the $\varphi$-types of the states in $C$. The
computation of the $\varphi$-types of the states in $M$ works
correctly, because we have a witness for every sentence in the type of some node in
our representation. This is by Lemma \ref{Lem:Models}, which ensures
that witnesses are in the node of the state or in one of its
children. Therefore, we cut below these witnesses when building the
finite representation. Consequently, the algorithm will accept.

On the other hand, if the algorithm accepts, it is straightforward to
construct a block tree model from the guessed representation.
The only critical point for soundness is the
verification of the $\varphi$-types guessed in Step 2, more precisely,
the computation of the $\varphi$-types of the states in $M$. First,
the $\varphi$-type of some state $s\in M$ contains only sentences that
hold at some state of the represented model below $s$. This can be assured
by looking only at states in $M$ and not at states in $C$. That the
$\varphi$-type of $s$ contains all sentences that hold below $s$ can be
assured by following the links represented by states in $C$.

The first two steps of the algorithm can be performed in exponential
time because the representation is of at most exponential size. That
Step 3 runs in exponential time follows from Theorem 4.5 of \cite{FdR05},
the truth of which is not affected by our modifications. The time
bounds for the other steps follow again from the exponential size
bound of the representation.
\end{Bew}

From Theorem \ref{ThAlg} and Corollary \ref{CorNEXPhard} we can
conclude Theorem \ref{ThHLDSAT}.

  \section{Decidability of Richer Hybrid 
              $\mathbf{\downarrow}$ Logics}
    \label{sec:undec}
    This section is concerned with satisfiability over transitive frames, transitive trees,
    and linear frames for extensions of $\mKlasse{HL}^\dna$.
    
    We investigate what happens if we extend the logic of the previous section with the $\at$-operator and/or
    the past modality $\Pas$. We will prove undecidability over transitive frames.
    
    Therefore, we will consider these logics over more restricted frame classes, 
    transitive trees and linear frames, where we will show them to be nonelementarily decidable.
    
    \subsection{Transitive Frames}
      \label{sec:trans}

      Over transitive frames, we cannot sustain decidability if we
      enrich $\mKlasse{HL}^\dna$ with $\at$ or the backward looking modality $\Pas$.
      We prove undecidability in both cases, making a detour via an undecidable fragment
      of first-order logic. The notation of such fragments is given in Section~\ref{sec:ML_HL}.

      We proceed in two steps.
      First, we show that
      $[\text{all},(4,1)]$\Probl{-trans-SAT} is undecidable. This is done by
      a reduction from $[\text{all},(0,1)]$\Probl{-SAT}.
      The undecidability of the latter
      is a consequence of the undecidability of contained traditional standard classes
      \cite{BGG97}. The second step consists of reductions from
      $[\text{all},(4,1)]$\Probl{-trans-SAT}
      to $\mKlasse{HL}^{\dnA,\at}$\Probl{-trans-SAT} and
      $\mKlasse{HL}^{\dnA}_{\Fut,\Pas}$\Probl{-trans-SAT}, respectively. To be more precise, the
      ranges of these reductions will be the fragments of the respective hybrid languages consisting
      of all nominal-free sentences.

      \begin{Lem}
        \label{lem:undecFOLtrans}
        $[\textup{all},(4,1)]$\Probl{-trans-SAT} is undecidable.
      \end{Lem}

      \begin{Bew}
        In order to obtain the required reduction from $[\text{all},(0,1)]$\Probl{-SAT}, we will
        transform a (not necessarily transitive)
        model satisfying $\alpha$ into a transitive one. Simply taking the transitive closure
        in most cases adds new pairs to the interpretation of the relation and is not sufficient for
        keeping the information which pairs were in the ``old'' relation and which pairs were not.
        This problem does not arise if we instead use a variation of  the \textit{zig-zag technique}
        successfully applied in~\cite{ABM00}
        for a reduction between a modal and a hybrid language.
        The core idea of this technique is to simulate
        an $R$-step $t_1Rt_2$ in the original model $\mKlasse{M} = (D,I)$ by a zig-zag transition in a model
        $\mKlasse{M}' = (D',I')$, where $I'(R)$ is transitive, as shown in Figure~\ref{fig:zigzag1}.
        \begin{figure}
          \parbox{.59\textwidth}{
            \begin{align*}
              (xRy)^t               & =  \EQ abc\,\big(xRa \und bRa \und bRc \und yRc                   \\
                                    & \!\!\!\!\und 0(x) \und 1(a) \und 2(b) \und 3(c) \und 0(y)\big),        \\
              (\neg\alpha)^t        & =  \neg(\alpha^t),                                             \\
              (\alpha \und \beta)^t & =  \alpha^t \und \beta^t,                                      \\
              (\EQ x\,\alpha)^t     & =  \EQ x\,\big(0(x) \und \alpha^t\big).
            \end{align*}
          }
          \hspace{-2mm}
          \parbox{.38\textwidth}{
            \includegraphics{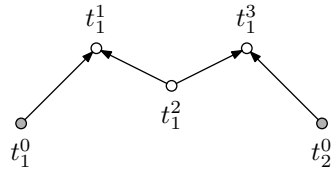}
          }
          \caption{The translation function and a zig-zag transition.}
          \label{fig:zigzag1}
        \end{figure}

        We define a translation function $(\cdot)^t$ using four extra predicate symbols $0,1,2,3$ according
        to Figure~\ref{fig:zigzag1}.
        The translation of the $xRy$-atoms exactly reflects the shown zig-zag transition.
        It is now straightforward to prove the following claim: \textit{For each formula $\alpha$, $\alpha$ is satisfiable
        iff $f(\alpha)$ is satisfiable in some model that interprets $R$ by a transitive relation.}
  
        Without loss of generality, we may assume that $\alpha$ has no free variables and that
        each variable is quantified exactly once. This can always be achieved by additional
        existential quantification and renaming, respectively.
  
        \bigskip\noindent
        \textbf{``$\mathbf{\Impl}$''.}~ Suppose $\alpha$ is satisfied by some model
        $\mKlasse{M} = (D,I)$. We construct a new model
        $\mKlasse{M}^4 = (D^4,I^4)$, where $D^4 = D^0\cup\dots\cup D^3$ using
        $D^i = \{d^i \mid d \in D\}$, for $i = 0,1,2,3$. The interpretation $I^4$ is defined by
        \begin{align*}
          I^4(R) & = \{(x^0, x^1),~(x^2,x^1),~(x^2,x^3),~(y^0,x^3) \mid (x,y) \in I(R)\}\quad\text{and} \\
          I^4(P) & = D^P, \quad P = 0,1,2,3.
        \end{align*}
        $I^4(R)$ codes an $I(R)$-transition from state $x$ to $y$ in \Klasse{M} as a sequence of
        backward and forward transitions from $x_0$ to $y_0$ via $x_1,x_2,x_3$ as shown in
        Figure~\ref{fig:zigzag1}. It is easy to see that $I^4(R)$ is transitive, since there is
        no state with incoming \textit{and} outgoing $I^4(R)$-edges.
  
        We now show that for all subformulae $\beta(x_1,\dots,x_m)$ of $\alpha$ and all
        $d_1,\dots,d_m \in D$: $\mKlasse{M} \wahr \beta\,[d_1,\dots,d_m]$ iff
        $\mKlasse{M}^4 \wahr \beta^t\,[d_1^0,\dots,d_m^0]$. This immediately implies that
        $\mKlasse{M}^4$ satisfies $\alpha^t$.
  
        We proceed by induction on $\beta$. The base case, $\beta = xRy$, is clear from the
        construction of $I^4(R)$. The Boolean cases are obvious. For the case
        $\beta = \EQ x\,\gamma$, we argue
        \begin{align*}
          \mKlasse{M} \wahr & ~\EQ x\,\gamma\,[d_1,\dots,d_m] \\
          & ~\Aequ~
            \EQ d \in D\, \big(\mKlasse{M} \wahr \gamma\,[d_1,\dots,d_m, x \mapsto d]\big) \\
          & ~\Aequ~
            \EQ d \in D\, \big(\mKlasse{M}^4 \wahr \gamma^t\,[d_1^0,\dots,d_m^0, x \mapsto d^0]\big) \\
          & ~\Aequ~
            \EQ d' \in D^4\, \big(\mKlasse{M}^4 \wahr (0(x) \und \gamma^t)\,[d_1^0,\dots,d_m^0, x \mapsto d']\big) \\
          & ~\Aequ~
            \mKlasse{M}^4 \wahr \EQ x\,(0(x) \und \gamma^t)\,[d_1^0,\dots,d_m^0] \\
          & ~\Aequ~
            \mKlasse{M}^4 \wahr (\EQ x\,\gamma)^t\,[d_1^0,\dots,d_m^0].
        \end{align*}
  
        The second and third line are equivalent due to the induction hypothesis. The equivalence of
        the third and fourth line follows from the definition of $R^4$; for the direction from
        below to above, one must take into account that because $x$ is interpreted by $d'$ and
        $0(x)$ is satisfied, $d'$ is indeed some $d^0$.
  
        \bigskip\noindent
        \textbf{\boldmath``$\lpmI$''.}~  Let $\mKlasse{M} = (D,I)$ be a model satisfying $\alpha^t$,
        where $I(R)$ is transitive.
        We construct a new model $\mKlasse{M}' = (D',I')$, where
        $D' = I(0)$ and
        \begin{align*}
          I'(R) =  \big\{(d,e) \in (D')^2  \mid \EQ abc \in D&\big((d,a), (b,a), (b,c), (e,c) \in I(R) \\
                                                             &\Und a \in I(1) \Und b \in I(2) \Und c \in I(3)\big)\big\}.
        \end{align*}
  
        We now show that for all subformulae $\beta(x_1,\dots,x_m)$ of $\alpha$ and all
        $d_1,\dots,d_m \in D$: $\mKlasse{M}' \wahr \beta\,[d_1,\dots,d_m]$ iff
        $\mKlasse{M} \wahr \beta^t\,[d_1,\dots,d_m]$. This immediately implies that
        $\mKlasse{M}'$ satisfies $\alpha$.
  
        Again, the proof is via induction on $\beta$. The base case, $\beta = xRy$, is clear from the
        construction of $I'(R)$ and the fact that the translation of $xRy$ requires $0(x)$ and $0(y)$.
        The Boolean cases are obvious. For the case $\beta = \EQ x\,\gamma$, we argue
        \begin{align*}
          \mKlasse{M}' \wahr & ~\EQ x\,\gamma\,[d_1,\dots,d_m] \\
          & ~\Aequ~
            \EQ d \in D'\, \big(\mKlasse{M}' \wahr \gamma\,[d_1,\dots,d_m, x \mapsto d]\big) \\
          & ~\Aequ~
            \EQ d \in I(0)\, \big(\mKlasse{M} \wahr \gamma^t\,[d_1,\dots,d_m, x \mapsto d]\big) \\
          & ~\Aequ~
            \EQ d \in D\, \big(\mKlasse{M} \wahr (0(x) \und \gamma^t)\,[d_1,\dots,d_m, x \mapsto d]\big) \\
          & ~\Aequ~
            \mKlasse{M} \wahr \EQ x\,(0(x) \und \gamma^t)\,[d_1,\dots,d_m] \\
          & ~\Aequ~
            \mKlasse{M} \wahr (\EQ x\,\gamma)^t\,[d_1,\dots,d_m].
        \end{align*}
  
        The induction hypothesis is applied between the second and third line. It is obvious that the
        third line implies the fourth; the backward direction is due to the fact that $x$ is interpreted
        by $d$ and $0(x)$ is satisfied, hence $d \in I(0)$.
  
        \bigskip\noindent
        This proves the above claim.
        Since $(\cdot)^t$ is an appropriate (even poly\-nomial-time) reduction function, we have
        established undecidability for $[\textup{all},(4,1)]$\Probl{-trans-SAT}.
      \end{Bew}
  
      \begin{Theo}
        \label{theo:dna_P_lb}
        \label{theo:dna_at_lb}
        $\mKlasse{HL}^{\dnA,\at}$\Probl{-trans-SAT} and
        $\mKlasse{HL}^{\dna}_{\Fut,\Pas}$\Probl{-trans-SAT} are undecidable.
      \end{Theo}
  
      \begin{Bew}
        We reduce $[\textup{all},(4,1)]$\Probl{-trans-SAT} to the two problems $\mKlasse{HL}^{\dnA,\at}$\Probl{-trans-SAT} and
        $\mKlasse{HL}^{\dna}_{\Fut,\Pas}$\Probl{-trans-SAT}, invoking a spy-point argument
        (for details of the spy-point technique see \cite{BS95,ABM99}). A spy-point is a state $s$ of a hybrid model that sees all other states
        and is named by a fresh nominal $i$. Since our reduction will not make use of any nominals,
        we can establish this undecidability result for the nominal-free fragments of the hybrid
        languages in question. We simply treat $i$ as a state variable and bind it
        to $s$.
  
        We first treat the case of $\mKlasse{HL}^{\dnA,\at}$ and define a
        translation function $(\cdot)^t$ from the first-order fragment to $\mKlasse{HL}^{\dnA,\at}$ by
        \begin{xalignat*}{2}
          (xRy)^t               & = \at_x\Diamond y,                      &
          (\neg\alpha)^t        & = \neg(\alpha^t),                       \\
          \big(P(x)\big){}^t    & = \at_x\,p,                             &
          (\alpha \und \beta)^t & = \alpha^t \und \beta^t,                \\
                                &                                         &
          (\EQ x\,\alpha)^t     & = \at_i\Diamond\!\!\dna x.\alpha^t.
        \end{xalignat*}
        The (polynomial) reduction function $f$ is defined by
        \[
            f(\alpha) = \Dna i.\big(\neg\Diamond i \und \Diamond \alpha^t\big).
        \]
  
        In order to argue that each formula $\alpha$ is satisfiable iff $f(\alpha)$ is satisfiable,
        we assume w.\,l.\,o.\,g.\ that $\alpha$ is a sentence (see proof of Lemma~\ref{lem:undecFOLtrans}).
        For the ``$\Impl$'' direction, suppose $\alpha$ is satisfied by a model
        $\mKlasse{M} = (D,I)$. By adding the spy-point $s$ to $D$, we obtain the hybrid model
        $\mKlasse{M}^h = (M^h, R^h, V^h)$, where $M^h = D \cup \{s\}$,
        $R^h = I(R) \cup \{(s,d) \mid d \in D\}$, and $V^h(p) = I(P)$. Clearly, $\mKlasse{M}^h$ satisfies
        $f(\alpha)$ at $s$\,---\,under any assignment, since $f(\alpha)$ is a sentence.
  
        For the ``$\lpmI$'' direction, suppose $f(\alpha)$ is satisfied at state $s$ of some hybrid model
        $\mKlasse{M} = (M,R,V)$. The composition of $f(\alpha)$ enforces $s$ to behave as the spy-point.
        It is easy to see that $\mKlasse{M}' = (M-\{s\},I)$, where $I(R) = R\einschr_{M-\{s\}}$ and
        $I(P) = V(p)$, satisfies $\alpha$.
  
        \medskip\noindent
        In the case of $\mKlasse{HL}^{\dna}_{\Fut,\Pas}$\,, we must simulate the \at\ operator using $\Pas$,
        which is possible in the presence of a spy-point and transitivity. We simply re-define $(\cdot)^t$
        by
        \begin{xalignat*}{2}
          (xRy)^t               & = \Pas\big(i \und \Fut(x \und \Fut y)\big), &
          (\neg\alpha)^t        & = \neg(\alpha^t),                           \\
          \big(P(x)\big){}^t    & = \Pas\big(i \und \Fut(x \und p)\big),      &
          (\alpha \und \beta)^t & = \alpha^t \und \beta^t,                    \\
                                &                                             &
          (\EQ x\,\alpha)^t     & = \Pas(i \und \Fut\!\!\dna x.\alpha^t)\,.
        \end{xalignat*}
        The rest of the proof is the same as
        for $\mKlasse{HL}^{\dnA,\at}$.
      \end{Bew}
  
    \subsection{Transitive Trees}
      \label{sec:tt}

      \newcommand{\Nat}{\ensuremath{(\mathbb{N},>)}}
      Over transitive trees, where decidability of $\mKlasse{HL}^{\dna}$
      is trivial, even the extension $\mKlasse{HL}^{\dnA,\at}_{\Fut,\Pas}$ is decidable. This is an immediate
      consequence of the decidability of the monadic second-order theory of the countably branching tree,
      S$\omega$S, \cite{BGG97}. However, we have to face a nonelementary lower bound in both cases
      $\mKlasse{HL}^{\dnA,\at}$ and $\mKlasse{HL}^{\dna}_{\Fut,\Pas}$. This is obtained by a reduction
      from the nonelementarily decidable $\mKlasse{HL}^{\dna}_{\Fut,\Pas}$\Probl{-\Nat-SAT}~\cite{FdRS03}.
      In the latter notation, \Nat\ stands for the frame class consisting only of the frame \Nat.

      \begin{Theo}
        \label{theo:dna_nonel}
        $\mKlasse{HL}^{\dna}_{\Fut,\Pas}$\Probl{-tt-SAT}, $\mKlasse{HL}^{\dnA,\at}$\Probl{-tt-SAT},
        and $\mKlasse{HL}^{\dnA,\at}_{\Fut,\Pas}$\Probl{-tt-SAT} are nonelementarily decidable.
      \end{Theo}
  
      \begin{Bew}
        Decidability immediately follows from decidability of S$\omega$S, using the Standard Translation
        $\ST$. For the nonelementary lower bound, we reduce $\mKlasse{HL}^{\dna}_{\Fut,\Pas}$\Probl{-\Nat-SAT}
        to $\mKlasse{HL}^{\dna}_{\Fut,\Pas}$\Probl{-tt-SAT} and $\mKlasse{HL}^{\dnA,\at}$\Probl{-tt-SAT},
        respectively.
  
        Let us first consider $\mKlasse{HL}^{\dna}_{\Fut,\Pas}$\Probl{-tt-SAT}. The frame \Nat\ is a special
        case of a transitive tree. Our language is strong enough to enforce that a transitive tree model is
        based on \Nat. All we have to do is to require two properties:
        \begin{Numm}
          \item Every point has at most one direct successor.
          \item The underlying frame is rooted.
        \end{Numm}
        Property (2) is expressed by $\Pas\Has\bot$.
        Property (1) can be formulated in the following way:
        For any point $x$, whenever $x$ has some successor, then we name one of the \textit{direct}
        successors $y$ and ensure that \textit{all} direct successors of $x$ satisfy $y$. This translates as
        \[
          \lambda = \Fut\top \impl \Fut^1\!\!\dna y. \Pas^1\Goi^1 y,
        \]
        where $\Fut^1$, $\Pas^1$, and $\Goi^1$ can be expressed by means of $\Unt$ and $\Sin$,
        for example $\Fut^1\varphi \equiv \Unt(\varphi,\bot)$. But $\Unt(\varphi,\psi)$
        can be simulated by $\dna x.\Fut\big(\varphi \und \Has(\Pas x \impl \psi)\big)$; analogously
        for $\Sin(\varphi,\psi)$.
  
        Hence $\lambda$ is expressible in our language and of constant length. An appropriate reduction function $f$
        is given by $f(\varphi) = \varphi \und \lambda \und \Has\lambda \und \Has\Goi\lambda \und \Pas\Has\bot$. It is easy to
        observe that $\varphi$ is satisfiable in some linear model iff $f(\varphi)$ is satisfiable in some
        transitive tree.

        \medskip\noindent
        In the case of $\mKlasse{HL}^{\dnA,\at}$\Probl{-tt-SAT}, we first have to simulate the $\Pas$ operator.
        This is done by a modified spy-point argument (for details of the spy-point technique see \cite{BS95,ABM99}).
        We simply label one point in the transitive tree by a fresh nominal $i$ and simulate each occurrence of
        $\Pas$ in $\varphi$ using $\dnA$, a fresh variable $v$, and $i$. This is done in the following translation function
        $(\cdot)^t : \mKlasse{HL}^{\dna}_{\Fut,\Pas} \to \mKlasse{HL}^{\dnA,\at}$.
        \begin{xalignat*}{2}
          a^t                    & = a, \qquad a \in \ATOM,      &
          (\Fut\psi)^t           & = \Diamond(\psi^t),           \\
          (\neg\psi)^t           & = \neg(\psi^t),               &
          (\Pas\psi)^t           & = \Dna v.\at_i\Diamond(\psi^t \und \Diamond v), \\
          (\psi_1 \und \psi_2)^t & = \psi_1^t \und \psi_2^t,     &
          (\dna x.\psi)^t        & = \Dna x.(\psi^t).
        \end{xalignat*}
        It is easy to see that for each model \Klasse{M} based on \Nat, for each point $x \in \mathbb{N}$,
        and for each formula $\varphi \in \mKlasse{HL}^{\dna}_{\Fut,\Pas}$:~
        whenever $i$ is true at the root $0$, then
        $\mKlasse{M},x \wahr \varphi ~\Aequ~ \mKlasse{M},x \wahr \varphi'$.

        The point $s$ labelled $i$ represents the root of the frame \Nat. In the language $\mKlasse{HL}^{\dnA,\at}$,
        it is not possible to express Property (2). This is in fact not necessary if we make sure that we never
        refer to the past of $s$ in our final translation of $\varphi$. Such a ``wrong'' reference can only
        appear when the \at\ operator is used in connection with nominals occurring in $\varphi$.
        Let $\NOM(\varphi) = \NOM \cap \Sub(\varphi)$, and let
        \[
            \textstyle
            \mu = \bigwedge_{j \in \NOM(\varphi)} \at_i\Diamond j.
        \]
        Now the formula $\dna i.(\Diamond\varphi^t \und \mu)$ does not contain any reference to any point before $s$.

        It remains to ensure Property (1). This is done by replacing
        $\lambda$ by
        \[
          \lambda' = \Diamond\top \impl\, \Dna x.\Diamond^1\!\!\dna y. \at_x\Box^1 y
        \]
        and again expressing $\Diamond^1$ and $\Box^1$ by means of $\Unt$, which can be simulated as shown in
        Section~\ref{sec:Intro}.
        Now, an appropriate reduction function is $f'$, where
        $f'(\varphi) = \Dna i.(\Diamond\varphi^t \und \mu \und \lambda' \und \Box\lambda')$.
      \end{Bew}

\subsection{Linear Frames}
  \label{sec:lin}

In the last part of this section we consider linear frames. A frame is called linear if it is irreflexive, transitive, and trichotomous. An important special case is the frame of the natural numbers with the usual ordering relation.

Hybrid $\downarrow$ languages over linear frames have already been addressed by Franceschet, de Rijke, and Schlingloff \cite{FdRS03}. They showed that satisfiability of $\mKlasse{HL}^{\dnA,\at}_{\Fut,\Pas}$ is nonelementary, even over natural numbers. This result also holds for $\mKlasse{HL}^{\dna}_{\Fut,\Pas}$, because $\at_i\varphi$ can be simulated by
$$\Pas(i\wedge\varphi)\vee(i\wedge\varphi)\vee\Fut(i\wedge\varphi).$$
While complexity drops down to NP for $\mKlasse{HL}^{\dnA}$, the last case, i.e. $\mKlasse{HL}^{\dnA,\at}$, was left open.
We will answer this question in the following.
\begin{Theo}\label{ThHLDSATlin}
The satisfiability problem for $\mKlasse{HL}^{\dnA,\at}$ over linear frames and over natural numbers is nonelementarily decidable.
\end{Theo}
\begin{Bew}
Only the lower bound has to be shown. We do so by giving a reduction from the satisfiability problem of first order logic over strings, a problem long known to have nonelementary complexity \cite{Sto74}.

Strings over a finite alphabet $\Sigma$ can be represented as 
$$(\{1,\ldots,n\},<,(P_{\sigma})_{\sigma\in\Sigma}),$$
 were $<$ is the usual ordering and $P_{\sigma}$ a unary relation for every $\sigma\in\Sigma$. As before, these structures can also be used for hybrid reasoning.

We give a translation $HT$ from first order logic into $\mKlasse{HL}^{\dnA,\at}$, such that for every string $\mKlasse{S}$ and every first order sentence $\varphi$,
$$\mKlasse{S}\models\varphi \iff \mKlasse{S'},s\models \Dna s.(HT(\varphi)\wedge \psi),$$
where $\mKlasse{S'}$ results from $\mKlasse{S}$ by adding a spy-point $s$ preceding all other states, what is ensured by $\psi$.
\begin{xalignat*}{2}
  \text{HT}(P_{\sigma}(x))     & ~=~ \at_s\Diamond(x\wedge p_{\sigma})     &
  \text{HT}(\neg\varphi)       & ~=~ \neg\text{HT}(\varphi)                  \\
  \text{HT}(x=y)               & ~=~ \at_s\Diamond(x\wedge y)              &
  \text{HT}(\varphi\wedge\psi) & ~=~ \text{HT}(\varphi)\wedge\text{HT}(\psi) \\
  \text{HT}(x<y)               & ~=~ \at_s\Diamond(x\wedge\Diamond y)    &
  \text{HT}(\exists x.\varphi) & ~=~ \at_s\Diamond(\dna x.\text{HT}(\varphi))
\end{xalignat*}

The \emph{only if} direction is obvious. For the \emph{if} direction, we have to state that $\mKlasse{S'}$ is a string and not just a linear frame. This is done by the formula
$$\psi = \textit{FL} \wedge \textit{DISCRETE} \wedge \textit{UNIQUE}.$$
The precise meaning of $\psi$ is that the subframe generated by $s$, but without the state $s$ itself, is a string.

There has to be at least one state in the string and there has to be a start and an end of the string, i.e., a state only preceded by $s$ and a state without successor.
$$\textit{FL} = (\Diamond\!\dna x.(\at_s\Box\neg\Diamond x)) \wedge (\Diamond\Box\bot)$$
We also have to ensure that the frame is not dense.
$$\textit{DISCRETE} = \Box(\Diamond\top\rightarrow(\dna x.\Diamond\!\dna y.\at_x\Box\Box\neg y))$$
Finally, every state has to carry a unique label from the finite alphabet.
$$\textit{UNIQUE} = \Box\bigvee_{\sigma\in\Sigma}(\sigma\wedge\bigwedge_{\sigma'\neq\sigma}\neg\sigma')$$
All other properties of a string are already ensured by linearity.
\end{Bew}

  \section{Hybrid Until/Since Logic over Transitive Frames and Transitive Trees}
    \label{sec:US-logics_trans}

    In this section, we will consider $\mKlasse{HL}^{\Exi}_{\Unt,\Sin}$\Probl{-trans-SAT}
    and $\mKlasse{HL}^{\Exi}_{\Unt,\Sin}$\Probl{-tt-SAT}.
    In~\cite{ABM00} it was shown that $\mKlasse{HL}^{\at}_{\Unt,\Sin}$\Probl{-SAT}
    is \compl{\EXP}.

    As for the lower bound, we establish a result as general as possible, namely \Hard{\EXP}
    of $\mKlasse{ML}_\Unt$\Probl{-trans-SAT} and $\mKlasse{ML}_\Unt$\Probl{-tt-SAT}.

    \begin{Theo}
      \label{theo:US_lb}
      $\mKlasse{ML}_\Unt$\Probl{-trans-SAT} and $\mKlasse{ML}_\Unt$\Probl{-tt-SAT} are \hard{\EXP}.
    \end{Theo}

    \begin{Bew}
      We will reduce the \textit{global} satisfiability problem for \Klasse{ML}
      to both our (local) problems $\mKlasse{ML}_{\Unt}\Probl{-trans-SAT}$ and
      $\mKlasse{ML}_{\Unt}\Probl{-tt-SAT}$ using the same reduction function. The
      global satisfiability problem is defined by
      \begin{gather*}
          \mKlasse{ML}\Probl{-GLOBSAT} =            \\[-3pt]
          \{\varphi \in \mKlasse{ML} \mid \varphi
          \text{~is true in \textit{all} states of some Kripke model \Klasse{M}}\}.
      \end{gather*}
      Its \Compl{\EXP} is a direct consequence
      of the \Compl{\EXP} of $\mKlasse{ML}^\Exi$-SAT
      \cite{Spa93a}.

      It may seem difficult to try reducing this problem over \textit{arbitrary} frames
      to our satisfiability problem over \textit{transitive} frames. The critical point lies
      in making a non-transitive model transitive: taking the
      transitive closure of its relation forces us to add new
      accessibilities that would disturb satisfaction of $\neg\Diamond$-formulae.
      Fortunately though, the $\Unt$ operator can make us distinguish the
      accessibilities in the original model from those that have been added to make
      the relation transitive. Hence, a translation of $\Diamond\varphi$ should demand:
      ``Make sure that the current state sees a state in which
      the translation of $\varphi$ holds, and that there is no state in between.'' This
      translates as $\Unt(\varphi^t,\bot)$ into the modal language.

      To construct the required reduction, we define a
      translation function $(\cdot)^t : \mKlasse{ML} \to \mKlasse{ML}_\Unt$ by
      \begin{xalignat*}{2}
        p^t                     & =  p,\qquad p \in \PROP,\qquad              &
        (\varphi \und \psi)^t   & =  \varphi^t \und \psi^t,                   \\
        (\neg\varphi)^t         & =  \neg(\varphi^t),                         &
        (\Diamond\varphi)^t     & =  \Unt(\varphi^t,\bot).
      \end{xalignat*}

      Using $(\cdot)^t$,  we construct a reduction function
      $f: \mKlasse{ML} \to \mKlasse{ML}_\Unt$  via $f(\varphi) = \varphi^t \und \Box\varphi^t$
      (which is clearly computable in polynomial time).
      It is straightforward to prove the following two claims for each $\varphi \in \mKlasse{ML}$.
      \begin{Numm}
        \item If $\varphi \in \mKlasse{ML}\Probl{-GLOBSAT}$, then $f(\varphi) \in \mKlasse{ML}_\Unt\Probl{-tt-SAT}$.
        \item If $f(\varphi) \in \mKlasse{ML}_\Unt\Probl{-trans-SAT}$, then $\varphi \in \mKlasse{ML}\Probl{-GLOBSAT}$.
      \end{Numm}

      Since each transitive tree is a transitive model, (1) and (2) imply the claim of this theorem.

      \medskip\noindent
      \textbf{(1).}~ Suppose $\varphi$ is satisfied in all states of some Kripke
      model $\mKlasse{M} = (M,R,V)$. By considering the submodel generated by some arbitrary
      state, we can assume w.\,l.\,o.\,g.\ that \Klasse{M} has a root $w_0$.

      Due to the \textit{tree model property} \cite{BdRV01} there exists a tree-like model
      (a model whose underlying frame is a tree) that satisfies $\varphi$ at all states, too.
      Hence we can suppose \Klasse{M} itself to be tree-like. From this model, we construct
      $\mKlasse{M}' = (M,R^+,V)$, which is clearly a transitive tree.

      Because of the tree-likeness of \Klasse{M}, we observe that for each pair $(w,v) \in R$,
      there exists no $u \in M$ \textit{between} $w$ and $v$ in terms of $R^+$, i.\,e.\
      no $u$ such that $wR^+u$ and $uR^+v$.
      By means of this observation, we show that for all states $m \in M$ and all formulae
      $\psi \in \mKlasse{ML}$:~ $\mKlasse{M},m \wahr \psi$ iff $\mKlasse{M}',m \wahr \psi^t$.
      This claim implies that $\mKlasse{M}',w_0 \wahr \varphi^t \und \Box\varphi^t$.
      It is proven by induction on the structure of $\psi$. The only interesting case is
      $\psi = \Diamond\vartheta$, and the necessary argument can be summarized as follows.
      \begin{align*}
        \mKlasse{M},m & \wahr \Diamond\vartheta \\
          & \Aequ~  \EQ n \in M (mRn \Und \mKlasse{M},n \wahr \vartheta)                \\
          & \Aequ~  \EQ n \in M (mRn \Und \mKlasse{M}',n \wahr \vartheta^t)            \displaybreak[0]\\
          & \Aequ~  \EQ n \in M \big(mR^+n \Und \mKlasse{M}',n \wahr \vartheta^t
                                      \Und \neg\EQ u\!\!\;\in\!\!\;M (mR^+u \Und uR^+n)\big)   \displaybreak[0]\\
          & \Aequ~ \mKlasse{M}',m \wahr \Unt(\vartheta^t,\bot)
      \end{align*}

      In this argument, the equivalence of the first and the second line follows from the induction
      hypothesis. The second and third line are equivalent due to the above observation.

      \medskip\noindent
      \textbf{(2).}~  Let $\mKlasse{M} = (M,R,V)$ be a transitive model
      and $w_0 \in M$ such that $\mKlasse{M},w_0 \wahr f(\varphi)$.
      Again, we restrict ourselves to the submodel generated by $w_0$. Hence
      all states of \Klasse{M} are accessible from $w_0$.

      Define a new Kripke model $\mKlasse{M}' = (M,R',V)$ from \Klasse{M}, where
      $R' =  \{(w,v) \in R \mid \neg\EQ u \in M (wRuRv)\}$.
      We show that for all states $m \in M$ and all formulae $\psi \in \mKlasse{ML}$:~
      $\mKlasse{M}',m \wahr \psi$ iff $\mKlasse{M},m \wahr \psi^t$.
      Again, we use induction on the structure of $\psi$ with the only interesting
      case $\psi = \Diamond\vartheta$ and the following argument.
      \begin{align*}
        & \mKlasse{M}',m \wahr \Diamond\vartheta \\
          & \Aequ~  \EQ n\!\in\!M (mR'n \Und \mKlasse{M}',n \wahr \vartheta)      \displaybreak[0]\\
          & \Aequ~  \EQ n\!\in\!M \big((n\!=\!w_0 \text{~or~} w_0Rn) \Und mRn
                    \Und\neg\EQ u(mRuRn) \Und \mKlasse{M},n \wahr \vartheta^t\big) \\
          & \Aequ~ \mKlasse{M},m \wahr \Unt(\vartheta^t,\bot)
      \end{align*}

      The equivalence of the first and the second line is due to the fact that \Klasse{M} is rooted,
      the definition of $R'$ as well as the induction hypothesis.
      Now, since $\mKlasse{M},w_0 \wahr
      \varphi^t \und \Box\varphi^t$, we conclude that
      for all states $x \in M$,~ $\mKlasse{M},x \wahr \varphi^t$. The previous claim implies
      that $\mKlasse{M}'$ satisfies $\varphi$ at all states.
    \end{Bew}

    \bigskip\noindent
    The upper bounds for $\mKlasse{HL}^{\Exi}_{\Unt,\Sin}$\Probl{-trans-SAT}
    and $\mKlasse{HL}^{\Exi}_{\Unt,\Sin}$\Probl{-tt-SAT} require separate treatment. As for
    $\mKlasse{HL}^{\Exi}_{\Unt,\Sin}$\Probl{-trans-SAT}, we use an embedding into an appropriate
    fragment of first-order logic. In order to eliminate transitivity, we ``simulate''
    this semantic property by syntactic means, namely using the operators $\Unt^{++}$ and $\Sin^{++}$
    defined in Section~\ref{sec:ML_HL}.

    \begin{Lem}
      \label{lem:equiv}
      For any $X \subseteq \{\at,\Exi\}$, the problems
      $\mKlasse{HL}^X_{\Unt,\Sin}\Probl{-trans-SAT}$ and
      $\mKlasse{HL}^X_{\Unt^{++},\Sin^{++}}\Probl{-SAT}$ are polynomially reducible
      to each other.
    \end{Lem}

    \begin{Bew}
      Either problem can be reduced to the other via a simple bijection
      $f : \mKlasse{HL}^X_{\Unt,\Sin} \to \mKlasse{HL}^X_{\Unt^{++},\Sin^{++}}$
      or its inverse, respectively. This function simply replaces every occurrence of $\Unt$
      (or $\Sin$, respectively) in the input formula by $\Unt^{++}$ (or
      $\Sin^{++}$, respectively). Obviously, $f$ and $f^{-1}$ can be computed in
      polynomial time. It is straightforward to inductively verify the following
      two propositions.
  
      \begin{Numm}
        \item For every $\varphi \in \mKlasse{HL}^X_{\Unt,\Sin}$\;: If $\varphi$ is satisfied
          in a state $m$ of some transitive model \Klasse{M}, then
          $\mKlasse{M},m \wahr f(\varphi)$.
        \item For all $\varphi \in \mKlasse{HL}^X_{\Unt^{++},\Sin^{++}}$\;: If $\varphi$ is
          satisfied in a state $m$ of some model $\mKlasse{M} = (M,R,V)$, then
          the transitive model $\mKlasse{M}' = (M,R^+,V)$ satisfies $f^{-1}(\varphi)$
          at $m$.
      \end{Numm}
      ~\par\vspace{-2\baselineskip}
    \end{Bew}

    \noindent
    Now it is not difficult anymore to obtain a \KK{\DEXP} upper bound for
    $\mKlasse{HL}^\at_{\Unt,\Sin}$\Probl{-trans-SAT} by an embedding into
    the loosely $\mu$-guarded fragment $\mu$LGF of first-order logic whose satisfiability problem
    is \compl{\DEXP}~\cite{GW99}.
    Only the $\Exi$ operator requires a more careful analysis.

    \begin{Theo}
      \label{theo:US_ub}
      $\mKlasse{HL}^\Exi_{\Unt,\Sin}$\Probl{-trans-SAT} is in \KK{\DEXP}.
    \end{Theo}

    \begin{Bew}
      We first embed $\mKlasse{HL}^\at_{\Unt^{++},\Sin^{++}}$ into the loosely
      $\mu$-guarded fragment $\mu$LGF of first-order logic \cite{GW99}. Since the satisfiability problem for $\mu$LGF-sentences
      is \compl{\DEXP}~\cite{GW99}, we obtain a \KK{\DEXP} upper bound for $\mKlasse{HL}^\at_{\Unt,\Sin}$\Probl{-trans-SAT}
      by Lemma~\ref{lem:equiv}.
      As a second step, we will show a reduction from $\mKlasse{HL}^\Exi_{\Unt,\Sin}$\Probl{-trans-SAT}
      to $\mKlasse{HL}^\at_{\Unt,\Sin}$\Probl{-trans-SAT}.

      For the embedding into $\mu$LGF, we enhance the Standard Translation $\ST$ (see Section~\ref{sec:ML_HL})
      by the rule
      \[
        \ST_x\big(\Unt^{++}(\varphi,\psi)\big)
                                    =  \EQ y\big[xR^+y \und \ST_y(\varphi)
                                        \und \AQ z\big((xR^+z \und zR^+y) \impl \ST_z(\psi)\big)\big]
      \]
      and an analogous rule for $\ST_x\big(\Sin^{++}(\varphi,\psi)\big)$.~ $\ST_y$ and $\ST_z$ are defined by
      exchanging $x,y,z$ cyclically.

      It remains to take care of the $R^+$ expressions. But $xR^+y$ can be expressed by
      \[
          \big[\LFP W(x,y). \big(xRy \oder \EQ z (zRy \und xWz)\big)\big]xy,
      \]
      yielding a $\mu$LGF-sentence with three variables.
      (If $\Unt^{++}$ operators are nested, variables can be ``recycled''.)
      The constants from the translations of nominals can be eliminated introducing new variables
      as shown in~\cite{Grae99}.
      The whole translation only requires time polynomial in the length of the
      input formula.

      \bigskip
      As for the more expressive language with $\Exi$, we can embed the stronger
      language $\mKlasse{HL}^\Exi_{\Unt,\Sin}$ into $\mKlasse{HL}^\at_{\Unt,\Sin}$ using a
      spy-point argument and exploiting the fact that we are restricted to transitive frames.
      A spy-point is a point $s$ that sees all other points and is named by the fresh nominal
      $i$. For details of this technique see \cite{BS95,ABM99}.

      By adding a spy-point to a transitive model, $\Exi\varphi$ can
      be simulated by $\at_i\Diamond\varphi$. Hence, if we take the translation
      $(\cdot)^t : \mKlasse{HL}^\Exi_{\Unt,\Sin} \to \mKlasse{HL}^\at_{\Unt,\Sin}$ that simply
      replaces all occurrences of $\Exi$ as shown, we obtain a reduction function
      $f : \mKlasse{HL}^\Exi_{\Unt,\Sin}\Probl{-trans-SAT} \to \mKlasse{HL}^\at_{\Unt,\Sin}\Probl{-trans-SAT}$
      by setting $f(\varphi) = i \und \neg\Diamond i \und \Diamond\varphi^t$.

      Clearly, $f$ is computable in polynomial time. It is straightforward to verify that $f$ is an
      appropriate reduction function: If $\varphi \in \mKlasse{HL}^\Exi_{\Unt,\Sin}$ is satisfied
      at some point of some transitive model, add the spy-point $s$ and the according accessibilities.
      The new model satisfies $f(\varphi)$ at $s$. For the converse, if a transitive model
      satisfies $f(\varphi)$ at some point $s$, then $s$ must be a spy-point, and $\varphi^t$ is
      satisfied at another point $m$. Remove $s$ and the according accessibilities from the model
      and observe that it now satisfies $\varphi$ at $m$.
    \end{Bew}

    \noindent
    A note on the discrepancy between the upper and lower bound for
    $\mKlasse{HL}^\Exi_{\Unt,\Sin}$\Probl{-trans-SAT}. Since the \KK{\DEXP} result for
    $\mu$LGF in~\cite{GW99} holds for sentences without constants only, constants\,---\,which
    arise from the translation of nominals\,---\, must be reformulated using new variables.
    This causes an unbounded number of variables in the first-order vocabulary, because we have no
    restriction on the number of nominals in our hybrid language.

    Could we assume that the number of nominals were bounded, then the described reduction
    would yield guarded fixpoint sentences of bounded width. In this case, satisfiability
    is \compl{\EXP}~\cite{GW99}. It is not known whether in the case of a bounded
    number of variables, but an arbitrary number of constants, satisfiability for
    $\mu$LGF-sentences also decreases from \KK{\DEXP} to \KK{\EXP}, as it is
    the case for the fragment without the $\mu$ operator~\cite{tCF04}. If there were a
    positive answer to this question, an \KK{\EXP} upper bound for our satisfiability
    problem would follow.

    \medskip\noindent
    We now show that
    $\mKlasse{HL}^{\Exi}_{\Unt,\Sin}$\Probl{-tt-SAT}
    is in \KK{\EXP}, using an embedding into \PDLtree, the propositional dynamic logic for sibling-ordered
    trees introduced in~\cite{Kra95,Kra97}. Finite, node-labelled, sibling-ordered trees
    are the logical abstraction of XML (eXtensible Markup Language) documents. In~\cite{ABD+05},
    it was shown that satisfiability of \PDLtree\ formulae at the root of finite trees
    (\Probl{\PDLtree-SAT}) is decidable in \KK{\EXP}.

    Since we are going to give an embedding into \PDLtree, we first introduce its syntax and semantics.
    \PDLtree\ is the language of propositional dynamic logic with four atomic programs
    \Left, \Right, \Up, and \Down\ that are associated with the relations ``left sister'',
    ``right sister'', ``parent'', and ``daughter'' in trees. It consists of all formulae
    of the form
    \[
        \varphi ::= p \mid \neg\varphi \mid \varphi\und\varphi' \mid \langle\pi\rangle\,\varphi\,,
    \]
    where $p \in \ATOM$ and $\pi$ is a program.
    Programs are defined by
    \[
        \pi ::= \Left \mid \Right \mid \Up \mid \Down \mid \pi;\pi' \mid \pi \cup \pi' \mid \pi^\ast \mid \varphi?\,,
    \]
    where $\varphi$ is a formula.
    We abbreviate $[\pi]\,\varphi := \neg\langle\pi\rangle\,\neg\varphi$ and $a^+ := a;a^*$ for atomic
    programs $a$.

    A \PDLtree\ model is a multi-modal model $\mKlasse{M} = (T, R_\Down, R_\Right, V)$, where $T$ is a finite
    tree with an order relation on all immediate successors of any node, $R_\Down$ is the successor
    relation and $R_\Right$ is the ``next-sister'' relation. The set of relations is extended to
    arbitrary programs as follows:
    \begin{xalignat*}{2}
      R_\Up             & = R_\Down^-\,,               &
      R_{\pi \cup \pi'} & = R_{\pi} \cup R_{\pi'}\,,   \\
      R_\Left           & = R_\Right^-\,,              &
      R_{\pi^\ast}      & = R_\pi^\ast\,,              \\
      R_{\pi;\pi'}      & = R_{\pi} \circ R_{\pi'}\,,  &
      R_{\varphi?}      & = \{(m,m) \mid \mKlasse{M},m \wahr \varphi\}\,.
    \end{xalignat*}

    The satisfaction relation for atomic formulae and Booleans is defined as for hybrid logic. The
    modal case is given by
    \[
      \mKlasse{M},m \wahr \langle\pi\rangle\,\varphi  \text{~~iff~~}
      \EQ n \in T (mR_\pi n \Und \mKlasse{M},n \wahr \varphi).
    \]

    A formula $\varphi$ is \textit{satisfiable} if and only if there exists a model
    $\mKlasse{M} = (T,R_\Down, R_\Right,V)$,
    such that $\mKlasse{M},m \wahr \varphi$, where $m$ is the root of $T$. For any $\varphi$, let
    $N(\varphi)$ be the set of all nominals occurring in $\varphi$.

    \begin{Theo}
      \label{theo:US_tt_ub}
      $\mKlasse{HL}^{\Exi}_{\Unt,\Sin}$\Probl{-tt-SAT} is in \KK{\EXP}.
    \end{Theo}

    \begin{Bew}
      We reduce $\mKlasse{HL}_{\Unt,\Sin}^{\Exi}$\Probl{-trans-SAT} to \Probl{\PDLtree-SAT}
      and define a translation
      $(\cdot)^t : \mKlasse{HL}_{\Unt,\Sin}^{\Exi} \to \text\PDLtree$ by
      \begin{xalignat*}{2}
        p^t                            & = p,\quad p \in \ATOM,                                          &
        (\Exi\varphi)^t                & = \langle\Up^\ast;\Down^\ast\rangle\,\varphi^t,                 \\
        (\neg\varphi)^t                & = \neg(\varphi^t),                                              &
        \big(\Unt(\varphi,\psi)\big)^t & = \big\langle(\Down;\psi^t?)^\ast;\Down\big\rangle\,\varphi^t,  \\
        (\varphi \und \psi)^t          & = \varphi^t \und \psi^t,                                        &
        \big(\Sin(\varphi,\psi)\big)^t & = \big\langle(\Up;\psi^t?)^\ast;\Up\big\rangle\,\varphi^t.
      \end{xalignat*}
      (nominals are translated into atomic propositions);

      Since \PDLtree\ has no nominals, we must enforce that (the translation of) each nominal is true at
      exactly one point by requiring
      \begin{align*}
          \nu(i) = \langle\Down^\ast\rangle\,i \;\und\; & [\Down^\ast]\Big(
                     i \impl \big([\Down^+]\,\neg i
                      \und  [\Up^+]\,\neg i    \\
                      & \und  [\Up^\ast;\Left^+;\Down^\ast]\,\neg i
                      \und  [\Up^\ast;\Right^+;\Down^\ast]\,\neg i
                  \big)\Big)
      \end{align*}
      to hold for each nominal $i$. As a reduction function, we have
      \[
          f(\varphi) = \langle\Down^\ast\rangle\,\varphi^t \und \textstyle\bigwedge_{i \in N(\varphi)} \nu(i).
      \]
  
      It is clear that $f$ is computable in polynomial time and straightforward to show that $f$
      is an appropriate reduction function: Suppose, $\varphi$ is satisfiable
      in some finite transitive tree model $\mKlasse{M} = (M,R,V)$ based on the tree $(M,R')$
      with root $w$. Then $f(\varphi)$ is satisfiable in $w$ of the \PDLtree\ model based on the tree $(M,R')$,
      equipped with the valuation $V$. For the converse, if $f(\varphi)$ is satisfied at the root of some
      \PDLtree\ model $\mKlasse{M} = (M, R_\Down, R_\Right, V)$, then $\varphi^t$ is true at some
      point $w$, and each nominal is true at exactly one point of \Klasse{M}. Hence $(M,R_\Down^+,V)$\,---\,where
      $R_\Down^+$ is the transitive closure of $R_\Down$\,---\,is a hybrid transitive tree model satisfying
      $\varphi$ at $w$.
  
      \medskip
      Now there is one drawback in the reduction via $f$. According to our definition of a tree,
      it is not necessary that a (transitive) tree is finite or has a root. A node can have infinitely
      many successors, or there may be an infinitely long forward or backward path from some point.
      For most practical applications these cases are certainly hardly of interest, but we
      strive for a more general result. If we do allow for infinite depth or
      width, the above translation into \PDLtree\,---\,which is interpreted over finite,
      rooted trees\,---\,is not sufficient.
  
      To overcome finiteness, it suffices to re-examine the proof for the \KK{\EXP} upper bound of
      \PDLtree-satisfiability in~\cite{ABD+05}. This proof in fact covers a more general result, too,
      namely that satisfiability of \PDLtree\ formulae over (not necessarily finite) trees is in
      \KK{\EXP}.
  
      To cater for the fact that ``our'' trees do not need to have roots, we first observe that
      satisfiability over \textit{rooted} transitive trees is reducible to satisfiability over
      (arbitrary) transitive trees, because a root is expressible by $\Pas\Has\bot$ in our language.
      Since the lower bound from Theorem~\ref{theo:US_lb} holds with respect to rooted transitive trees,
      it also holds for arbitrary ones.
  
      In order to obtain the upper bound with respect to arbitrary transitive trees, we propose a
      modification of the above reduction via $f$. The basic idea is to turn the backward path from the
      node $w$ (that is to satisfy $\varphi$) into a forward path, such that $w$ becomes the root
      of the transformed model. Thus all predecessors of $w$ (and \textit{their} predecessors)
      become successors and must be marked by a fresh proposition $\flat$. (See Figure~\ref{fig:predsucc}.)

      \begin{figure}
        \begin{center}
          \includegraphics{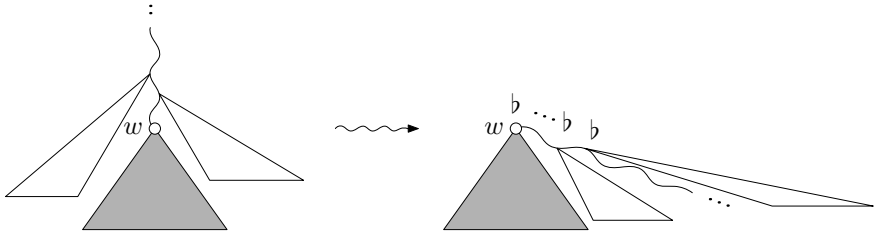}
        \end{center}
        \caption{Making predecessors successors.}
        \label{fig:predsucc}
      \end{figure}

      As a first step, we construct a new translation $(\cdot)^{t\flat}$ from $(\cdot)^t$ retaining
      all but the $\Unt$/$\Sin$-cases. For $\Unt$/$\Sin$, we replace all occurrences of the
      programs $\Down$ and $\Up$ by programs that incorporate the new structure and the fact that
      for $\flat$-nodes, their predecessors used to be their successors, and their $\flat$-successors
      used to be their predecessors. We define
      \begin{align*}
        \big(\Unt(\varphi,\psi)\big)^{t\flat}
             & = \big\langle(\Dn';\psi^{t\flat}?)^\ast;\Dn'\big\rangle\,\varphi^{t\flat} \qquad\text{and} \\
        \big(\Sin(\varphi,\psi)\big)^{t\flat}
             & = \big\langle(\Up';\psi^{t\flat}?)^\ast;\Up'\big\rangle\,\varphi^{t\flat}, \qquad\text{where}
      \end{align*}
      \[
        \Dn' = \big(\Down;\neg\flat?\big) \cup (\flat?;\Up) \quad\text{and}\quad
        \Up' = (\neg\flat?;\Up) \cup (\flat?;\Down;\flat?)\,.
      \]

      Note that we do not change the translation of $\Exi\varphi$. The only thing that remains to do
      is to enforce that there is exactly one path at whose every node $\flat$ is true. This means that
      $\flat$ must be true at the root node and at exactly one successor of each node
      satisfying $\flat$. This can be expressed by
      \begin{align*}
          \beta = \flat
                  & \und
                  [\Down^\ast]\,\Big(\flat \impl \big([\Left^+]\,\neg\flat \und [\Right^+]\,\neg\flat
                                                      \und \langle\Down\rangle\,\flat\big)\Big)\\
                  & \und
                  [\Down^\ast]\,\big(\neg\flat \impl [\Down]\neg\flat\big)\,.
      \end{align*}
      It is now straightforward to show that $f^\flat$, given by
      $
          f^\flat(\varphi) = \varphi^{t\flat} \und \beta \und \textstyle\bigwedge_{i \in N(\varphi)} \nu(i)\,,
      $
      is indeed an appropriate reduction function.

      (Note that $\varphi^{t\flat}$ replaces $\langle\Down^\ast\rangle\,\varphi^{t\flat}$,
      because we have turned $w$ into the new root node.)
    \end{Bew}

  \section{Conclusion}
    \label{Conclusion}

    We have established two groups of complexity results for hybrid logics over three
    temporally relevant frame classes: transitive frames, transitive trees, and linear frames.

    First, we have ``tamed'' $\mKlasse{HL}^{\dnA}$ over transitive frames showing that
    $\mKlasse{HL}^{\dnA}$\Probl{-trans-SAT} is \compl{\NEXP}. The key step of our proof was to
    find a finite representation of transitive models for this logic.
    In contrast, we proved that $\mKlasse{HL}^{\dnA,\at}$\Probl{-trans-SAT} and
    $\mKlasse{HL}^{\dnA}_{\Fut,\Pas}$\Probl{-trans-SAT} are undecidable.
    In this context, the question arises whether the \textit{multi-modal} variant of
    $\mKlasse{HL}^{\dnA}$ over transitive frames is still decidable.

    Over transitive trees, we showed three enrichments of $\mKlasse{HL}^{\dnA}$ to be decidable,
    albeit nonelementarily, namely $\mKlasse{HL}^{\dnA,\at}$\Probl{-tt-SAT},
    $\mKlasse{HL}^{\dnA}_{\Fut,\Pas}$\Probl{-tt-SAT}, and
    $\mKlasse{HL}^{\dnA,\at}_{\Fut,\Pas}$\Probl{-tt-SAT}.
    Concerning linear frames, we obtained the same result for 
    $\mKlasse{HL}^{\dnA,\at}$\Probl{-lin-SAT}, an issue left open in \cite{FdRS03}.

    In the third part of our work, we established an \KK{\EXP} lower bound for
    $\mKlasse{ML}_{\Unt}$\Probl{-trans-SAT} and
    $\mKlasse{ML}_{\Unt}$\Probl{-tt-SAT} and matched the latter with an \KK{\EXP}
    upper bound for $\mKlasse{HL}^\Exi_{\Unt,\Sin}$\Probl{-tt-SAT}. This is the same complexity as
    for satisfiability over \textit{arbitrary} frames for the same language.
    As for $\mKlasse{HL}^\Exi_{\Unt,\Sin}$\Probl{-trans-SAT}, we have given a
    \KK{\DEXP} upper bound. We conjecture \Compl{\EXP}.

    Over linear frames, the complexity of hybrid $\Unt$/$\Sin$ logic is still 
    open. 
    As a special case, satisfiability of $\mKlasse{HL}^\at_{\Unt,\Sin}$
    over $(\mathbb{N},>)$ is known to be \compl{PSPACE} \cite{FdRS03}. Moreover,
    in \cite{Rey03} it was shown that $\mKlasse{ML}_\Unt$ is \compl{PSPACE} with respect to
    general linear time. Over this frame class\,---\,which does not properly contain that of
    linear orders\,---\,the temporal hybrid languages from this paper could be
    re-examined in future work.

  \section*{Acknowledgments}

    We thank Balder ten Cate, Massimo Franceschet, and Maarten Marx for
    helpful suggestions, discussions, and comments.

  \bibliographystyle{plain}

\end{document}